\documentclass[journal]{IEEEtran}

\usepackage{ifpdf}
\usepackage{subfigure}
\usepackage{graphicx}
\DeclareGraphicsExtensions{.pdf,.jpg,.png}
\usepackage[cmex10]{amsmath}
\usepackage{amssymb}
\usepackage{array}
\usepackage{mdwmath}
\usepackage{mdwtab}
\usepackage{pslatex}
\usepackage{xcolor}
\usepackage{textcomp}
\usepackage{placeins}
\usepackage{eqparbox}
\usepackage{url}
\usepackage{stfloats}
\usepackage{multicol}
\usepackage{subfigure}
\usepackage{float}
\usepackage{algorithmicx}
\usepackage{algpseudocode}
\usepackage{amsthm}
\usepackage{comment}
\usepackage{cite}
\usepackage{color, soul}
\usepackage{enumerate}
\usepackage{bbm}
\usepackage{threeparttable}
\usepackage{booktabs}
\usepackage{array}
\usepackage{balance}
\usepackage{color}

\newtheorem{thm}{Theorem}

\begin{document}

%\title{Capacity of Joint Terahertz and Millimeter Wave Systems with multi-connectivity Operation}

\title{User Association and Multi-connectivity Strategies in Joint Terahertz and Millimeter Wave 6G Systems\vspace{-0mm}}
	
\author{Eduard~Sopin, Dmitri~Moltchanov, Anastasia~Daraseliya, Yevgeni~Koucheryavy, and Yuliya Gaidamaka\vspace{-3mm}

%\thanks{The authors are with the Faculty of Information Technology and Communication Sciences, Department of Electrical Engineering, Tampere University, Finland. Email: \{firstname.lastname@tuni.fi\}.}
\thanks{D. Moltchanov and Y. Koucheryavy are with Tampere University, Finland. Email:~{dmitri.moltchanov}@tuni.fi, {yevgeni.koucheryavy}@tuni.fi}
\thanks{ E. Sopin, A. Daraseliya, and Yu. Gaidamaka are independent researchers.}
}

% E. Sopin, A. Daraseliya, and Yu. Gaidamaka are with Peoples' Friendship University of Russia (RUDN University), Moscow, Russia
% Email:~{sopin-es,daraselia-av,gaydamaka-yuv}@rudn.ru.}
%\thanks{D. Moltchanov and Y. Koucheryavy are with Tampere University, Finland. Email:~{dmitri.moltchanov}@tuni.fi, {yevgeni.koucheryavy}@tuni.fi}
%\thanks{E. Sopin and Yu. Gaidamaka are also with Institute of Informatics Problems, Federal Research Center Computer Science and Control of Russian Academy of Sciences, Moscow, Russia.}

\maketitle

\begin{abstract}
Terahertz (THz) wireless access is considered as a next step towards sixth generation (6G) cellular systems. By utilizing even higher frequency bands than 5G millimeter wave (mmWave) New Radio (NR), they will operate over extreme bandwidth delivering unprecedented rates at the access interface. However, by relying upon pencil-wide beams, these systems will not only inherit mmWave propagation challenges such as blockage phenomenon but introduce their own issues associated with micromobility of user equipment (UE). In this paper, we analyze and compare user association schemes and multi-connectivity strategies for joint 6G THz/mmWave deployments. Differently, from stochastic geometry studies, we develop a unified analytically tractable framework that simultaneously accounts for specifics of THz and mmWave radio part design and traffic service specifics at mmWave and THz base stations (BS). Our results show that (i) for negligible blockers density, $\lambda_B\leq{}0.1$ bl./$m^2$, the operator needs to enlarge the coverage of THz BS by accepting sessions that experience outage in case of blockage (ii) for $\lambda_B>0.1$ bl./$m^2$, only those sessions that does not experience outage in case of blockage need to be accepted at THz BS, (iii) THz/mmWave multi-connectivity improves the ongoing session loss probability by $0.1-0.4$ depending on the system parameters.
\end{abstract}
	

% To this aim, we develop a unified analytically tractable framework that accounts for specifics of THz and mmWave propagation conditions including blockage and micromobility. Differently from stochastic geometry studies, the comparison is based not only on radio channel characteristics but explicitly accounts for traffic service specifics at mmWave and THz BSs.

%------------------------------------------------

\section{Introduction}\label{sec:intro}

% The role of THz systems in cellular landscape

Similarly to millimeter wave (mmWave) New Radio (NR) systems providing capacity boost for cellular infrastructure in places with high traffic demands \cite{shafi2020real}, terahertz (THz) access systems are expected to be utilized in locations with the need for extraordinary data rates \cite{giordani2020toward}. However, in addition to extreme capacity, these systems bring several unique challenges to system designers, making efficient deployment and utilization of such systems a complex task.

% Link layer

The propagation and link layer specifics of THz communications have been studied fairly well so far. Similarly to mmWave systems, THz communications are subject to blockage phenomenon \cite{bilgin2019human}. In addition to higher free-space propagation losses, THz propagation is also much heavier affected by atmospheric absorption as compared to mmWave band \cite{jornet2011channel}. Finally, by relying upon extremely directional antenna radiation patterns with half-power beamwidth (HPBW) of the main lobe approaching $1^\circ$, these systems may suffer from frequent outages due to UE micromobility \cite{petrov2020capacity,stepanov2021statistical}. These specifics make THz links highly unreliable requiring natural support from other technologies to ensure session service continuity at the THz air interface.

% Intra-RAT multi-connectivity

Future THz wireless access systems will target the support of rate-greedy applications such as virtual/augmented reality (VR), 8/16K streaming, holographic telepresence that are inherently  sensitive to outages \cite{vannithamby2017towards}. To improve session continuity of applications running over inherently unreliable wireless technologies, 3GPP has recently proposed the multi-connectivity functionality \cite{3gpp_MC}. According to it, UE is allowed to maintain two or more active connections to nearby base stations (BS) of the same or different radio access technologies (RAT), referred to as inter- and intra-RAT multi-connectivity, respectively. The intra-RAT multi-connectivity has been studied in detail in context of mmWave 5G NR systems \cite{polese_dual_con,tesema_gc_multicon,margarita2018multicon,begishev2021joint}, where it was shown to drastically improve outage performance of UEs. However, the limited coverage of prospective THz BSs as compared to mmWave NR BSs will require very dense deployments to efficiently utilize intra-RAT multi-connectivity. 

% Inter-RAT multi-connectivity

%The inter-RAT multi-connectivity can be utilized as one of the options to address inherently unreliable performance of THz links. 

%The inter-RAT multi-connectivity has also been studied in context of mmWave 5G NR and microwave LTE/NR/WiFi systems. Specifically, the authors in \cite{begishev2021performance} considered joint LTE and mmWave NR deployment and have shown that it may improve session continuity. Similar results have been reported in \cite{moltchanov2021performance,petrov2018achieving} for different deployment conditions. 

% Thus, the design of inter-RAT multi-connectivity to improve session continuity in future 6G THz/mmWave deployments is a difficult task.

To improve reliability of user sessions in future 6G THz/mmWave deployments, inter-RAT multi-connectivity operation can be utilized. However, the only other RAT having comparable albeit much smaller resources at the air interface is 5G NR operating in mmWave band. In spite both technologies target similar type of rate-greedy outage-sensitive applications, there is an inherent capacity mismatch between technologies. Thus, one needs to account not only for the radio specifics but for details of the resource allocation process at BSs. Furthermore, differently from the joint usage of microwave ($\mu$Wave) and mmWave technologies \cite{moltchanov2021performance,begishev2021performance}, both mmWave and THz RATs are subject to blockage events that require careful design of user association strategies. Finally, THz RAT is also affected by UE micromobility. To the best of the authors' knowledge, there are no studies assessing the use of inter-RAT multi-connectivity in 6G THz/mmWave deployments.

% System-level is not well studies so far

%So far there were just few attempts to assess system-level integration of THz systems into the future wireless landscape. The early attempts in the field, e.g., \cite{petrov2016applicability}, demonstrated that the use of THz hotspots may indeed provide a giant capacity boost for applications with caching capabilities drastically offloading LTE infrastructure. The recent wave of studies spawned by the completion of 5G NR standardization efforts, e.g., \cite{yang20196g,huang2019survey,polese2020toward} further established the role of THz access as a vital part of 6G systems. In context of inter-RAT multi-connectivity operation in THz systems the problem of optimal user associations with RATs becomes critical. The only study published so far, where user associations in THz systems have been considered is due to Hassan et al. \cite{hassan2020user}. However, the authors mainly concentrated on system-level metrics proposing schemes that minimized the standard deviation of the network load in inherently heterogeneous systems with LTE, NR and THz access technologies.

% What we do

In this paper, we fill the abovementioned gap by evaluating and comparing performance of user association schemes and multi-connectivity strategies in joint THz/mmWave radio access networks. Contrarily to the previous studies of user associations in heterogeneous access systems, we explicitly account for not only radio part specifics of considered technologies including propagation, dynamic blockage and micromobility effects but the traffic service process at mmWave and THz BSs. The performance of the considered user association schemes and multi-connectivity strategies are assessed and compared based on the user- and system-level metrics including new session loss capabilities, ongoing session loss probability as well as system resource utilization. 

% Main contributions

The main contributions of our study are:
\begin{itemize}
    \item{a mathematical framework based on stochastic geometry and queuing theory allowing for unified characterization of association schemes and multi-connectivity strategies by accounting for both radio and service part specifics;}
    \item{numerical analysis of user association schemes and multi-connectivity strategies for THz/mmWave deployments;}
    \item{numerical results showing that: (i) there is trade-off between new and ongoing session loss probabilities that depends on the choice of association and multi-connectivity strategy, (ii) the choice of the optimal association scheme mainly depends on the blockers density, (iii) tolerance of applications to short-term outages caused by antenna misalignment may greatly improve user performance, and (iv) intra-RAT multi-connectivity improves the ongoing session loss probability by approximately $0.1-0.4$.}
\end{itemize}

\begin{table} [!t]
	\centering
	\caption{Notation utilized in the paper.}
	\label{table:notation}
	\begin{tabular}{p{0.22\columnwidth}p{0.68\columnwidth}}
		\hline 
		\textbf{Notation}& \textbf{Description}\\ 
		\hline 
		\hline
		$f_{M,c},f_{T,c}$ & mmWave and THz carrier frequencies, Hz\\
		\hline
		$B_{M},B_{T}$ & mmWave and THz BS bandwidths, Hz\\
		\hline
		$C$& requested rate of sessions, Mbps\\ 
		\hline 
		$\lambda_A$ & intensity of session arrivals, sess./s/m$^2$\\
		\hline
		%$\Lambda$ & overall session arrivals , sess./s\\
		%\hline
		$K$ & number of THz BSs in the coverage of mmWave BS\\
		\hline 
		$\lambda_B$ & density of blockers, bl./m$^2$\\
		\hline
		$r_M$& mmWave BSs coverage radius, m \\
		\hline 
		$r_T$ & THz BS coverage radius, m  \\
		\hline
		$h_M$& mmWave BS height, m \\ 
		\hline
%		$v$ & blocker speed\\
%		\hline
		$v_B,h_B,r_B$ & blocker speed (m/s), height (m), and radius (m)\\
		\hline 
%		$r_B$ & blocker radius \\ 
%		\hline 
	    $h_U$ & UE height, m\\
	    \hline 
		$P_M$ & mmWave BS emitted power, W\\
		\hline
		$N_0$ & thermal noise power, dBi\\
		\hline
		$G_{M,B}$ & mmWave BS antenna array gain \\
		\hline
		$G_{M,U}$ & mmWave UE antenna array gain \\
		\hline
	    $A_{M}$ & mmWave propagation coefficient\\
		\hline
		$\zeta_{M,1},\zeta_{M,2}$ &  mmWave path loss exponents\\
		\hline
		$M_{M,2}$ & mmWave shadow fading in blocked state, dB\\
		\hline
		$p_{M,O},p_{T,O}$ & cell edge outage probability at mmWave/THz BS \\
		\hline
		$S_{M,min}$ & SINR outage threshold, dB\\
		\hline
		$\sigma_{M,2}$ & STD of the shadow fading in LoS blocked state, dB\\
		\hline
		$h_{M,B}$ & mmWave BS height, m\\
		\hline
		$P_T$ & THz BS emitted power, W\\
		\hline
		$G_{T,B}$ & THz BS antenna array gain \\
		\hline
		$G_{T,U}$ & THz UE antenna array gain \\
		\hline
	    $A_{T}$ & THz propagation coefficient\\
		\hline
		$\zeta_{T,1},\zeta_{T,2}$ &  THz path loss exponents\\
		\hline
	    $M_{M,I},M_{T,I}$ & mmWave and THz interference margins, dB\\
		\hline
		$L_{A}(f_{T,c},r)$ & absorption loss, dB\\
	    \hline 
	    $K(f)$ & absorption  coefficient, m$^{-1}$\\
		\hline
		$h_{T,B}$ & THz BS height, m\\
		\hline
		$N_{T,B,V}\times{}N_{T,B,H}$ & THz BS antenna configuration, el.$\times$el.\\
		\hline
		$N_{T,U,V}\times{}N_{T,U,H}$ & THz UE antenna configuration, el.$\times$el.\\
		\hline
		$N_{M,B,V}\times{}N_{M,B,H}$ & mmWave BS antenna configuration, el.$\times$el.\\
		\hline
		$N_{M,U,V}\times{}N_{M,U,H}$ & mmWave UE antenna configuration, el.$\times$el.\\
		\hline
		$\delta$ & THz array switching time, s\\
		\hline
		$T_B$ & THz beamalignment duration, s \\
		\hline
		$T_O$ & outage tolerance time of applications, s\\
		\hline
		$\mu^{-1}$ & mean service time of sessions, s\\
		\hline
		$p_{1,r},p_{2,r}$ & session resource requirements pmfs\\
		\hline
		$\nu$ & blockage intensity at mmWave BS, bl./s \\
		\hline
		$\nu_{M}$ & outage intensity caused by micromobility events/s \\
		\hline
		$\nu_{B}$ & outage intensity caused by blockage, events/s\\
		\hline
		$p_{0,1},p_{0,2}$ & outage fraction for on-demand/periodic alignment \\
		\hline
		$T_{O,1},T_{O,2}$ & outage for on-demand/periodic beamalignment, s\\
		\hline
		$T_A$ & time to outage due to micrimobility, s\\
		\hline
		$f_{T_A}(t)$ & pdf of time to outage in presence of micromobility \\
		\hline
		$T_U$ & periodic beamalignment interval, s\\
		\hline
		$\beta_B$ & connection recovery intensity for blockage, events/s\\
		%average outage time caused by blockage\\
		\hline
		$\beta_M$ & connection recovery intensity for micromob., events/s\\
		%average outage time caused by beamalignment\\
		\hline
		$\epsilon_j$ & probability that the UE is assigned the MCS $j$  \\
		\hline
		$p_T$ & THz UE association probability\\
		\hline
		$x(t), y(t)$ & random displacements processes over $x$- and $y$-axis\\
		\hline
		$\phi(t),\theta(t)$ & yaw and pitch rotational mobility processes\\
		\hline
		$\alpha$ & antenna array HPBW, $^\circ$\\
		\hline
		$\theta_m,\theta_{3db}^{\pm}$ & array maximum and $\pm$ 3 dB points, $^\circ$\\
		\hline
%		$\theta_{3db}^{\pm}$ & upper and lower 3-dB  points\\
%		\hline
		$p_B(y)$ & blockage probability at distance $y$\\
		\hline
		$J$ & number of NR MCSs\\
		\hline
	\end{tabular} 
	\vspace{-4mm}
    \end{table}

% How it is all organized

Our paper is organized as follows. In Section \ref{sec:syst} we introduce our system model. Performance evaluation framework is developed in Section \ref{sec:perf} and parameterized using radio part parameters in Section \ref{sec:param}. Numerical analysis is conducted in Section \ref{sec:num}. Conclusions are drawn in the last section. 

%--------------------------------------------------

\section{Related Work}\label{sec:rel}

% THz alone

System-level performance analysis of future 6G THz systems has been mainly studied so far by utilizing stochastic geometry approach. The studies mainly concentrated on characterizing impairments produced by blockage and micromobility as well as on the techniques utilized to mitigate them. Specifically, the authors in \cite{petrov2017interference,kokkoniemi2017stochastic} characterized interference and signal-to-interference plus noise ratio (SINR) at UEs by accounting for blockers density and directional antennas. Further, studies extending those works to deployment specifics have been published, see Shafie et al. \cite{shafie2021coverage} for 3D, Wu et al. \cite{wu2020interference} for indoor deployments, etc. Recently, the authors also investigated the effect of 3GPP multiconnectivity in dynamic blockage environment in \cite{shafie2020multiconnectivity}. These results have been further extended to account for micromobility in \cite{moltchanov2021ergodic}. In both studies, the authors concentrate on capacity and outage probability showing that multiconnectivity improves both metrics.

% Joint systems

Much less is known about system-level performance of joint operation of mmWave/THz systems. Among few others, is the study in \cite{wang2021joint}, where the authors formulated the optimization problem of determining user associations in joint mmWave/THz systems to improve UE throughput in these systems. Another similar study targeting associations in these systems is \cite{hassan2020user}, where a coexisting $\mu$Wave and THz system has been considered. By utilizing system-level simulations heuristic UE associations algorithms have been proposed. However, similarly to THz-only systems, the abovementioned studies mainly concentrate on elastic traffic patterns.

% Why we need new techniques

The main rationale for utilizing mmWave and THz systems jointly is the type of traffic they target. By providing extreme amount of resources at the air interface both RATs target rate-greedy non-elastic applications such as AR/VR, 8/16K streaming, holographic telepresence \cite{vannithamby2017towards}. One of the inherent properties of joint mmWave/THz systems is that both of them are subject to dynamic blockage while THz systems may further be affected by micromobility. There are few studies published to date targeting non-elastic traffic in mmWave only systems \cite{begishev2021joint} or mmWave/$\mu$Wave systems \cite{begishev2021performance} with multiconnectivity support. Among other conclusions, \cite{begishev2021performance} shows that temporal offloading of rate-greedy connections to $\mu$Wave systems leads to detrimental effects in terms of UE having $\mu$Wave interface only, while \cite{begishev2021joint} illustrates that mmWave system alone may still lead to non-negligible loss probability of sessions accepted to service.

The conventional analysis tool utilized in the past -- stochastic geometry -- is no longer sufficient alone for mmWave/THz systems as it inherently assumes elastic traffic. To make conclusions on the performance characteristics of the session service process in joint mmWave/THz systems we need to take into account not only the specifics of the radio part and stochastic factors related to the randomness of channel and UE locations, but the traffic service dynamics at BSs by joining the tools of stochastic geometry and queuing theory.

%Thus, the performance evaluation frameworks tailored towards mmWave and THz RATs have to take into account not only the specifics of the radio part and stochastic factors related to the randomness of UE locations, but the traffic service dynamics at BSs by joining the tools of stochastic geometry and queuing theory.

% are expected to primarily target applications generating non-elastic/adaptive traffic and requiring high and guaranteed bitrates at the air interface and having no or limited application layer adaptmiation capabilities

%--------------------------------------------------

\section{System Model}\label{sec:syst}

In this section, we introduce our system model. We define the deployment, dynamic blockage, propagation, antenna, beamalignment, and micromobility models, and then complement them with the resource allocation and traffic assumptions. Finally, we introduce the user associations and dynamic multi-connectivity assumptions as well as the metrics of interest. The notation used in this paper is provided in Table \ref{table:notation}.

\subsection{Deployment Model}

% mmWave deployment

\begin{figure*}[!t]
    \centering
    \includegraphics[page=1,clip,trim=20cm 19cm 20cm 10.2cm, width=1.\textwidth]{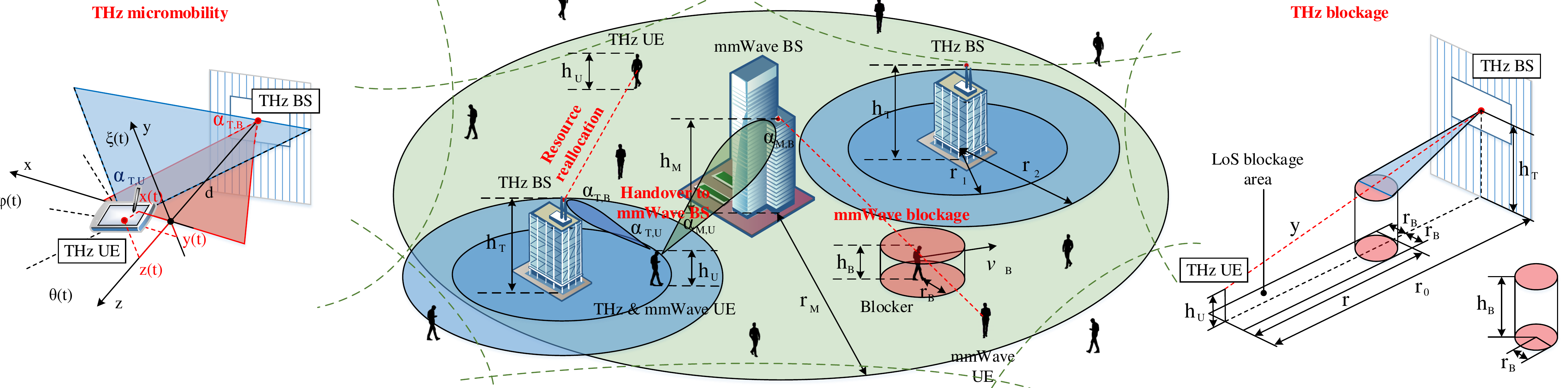}
    \caption{The considered 6G deployment with joint THz/mmWave BSs.}
    \label{fig:main}
    \vspace{-3mm}
\end{figure*}

We assume a mature phase of mmWave NR market penetration by considering a well-provisioned mmWave BS deployment. We tag and consider an arbitrarily chosen circularly-shaped service area of mmWave BS with radius $r_M$, where $r_M$ is such that no outage happens at the cell boundary in line-of-signt (LoS) blocked state, see Fig. \ref{fig:main}. This radius is determined in Section \ref{sec:param} by utilizing the set of NR modulation and coding schemes (MCS, \cite{nrmcs}) and propagation model defined below.

% THz deployment

In the coverage area of mmWave BS an operator is assumed to deploy $K$ THz BSs. The geometric locations of these THz BSs are uniformly distributed. The maximal feasible coverage radius of these THz BSs is determined by the propagation model introduced in what follows, while the actual coverage $r_T$ depends on the user associations schemes discussed below. The heights of mmWave and THz BSs are constant and given by $h_M$ and $h_T$. The carrier frequencies are denoted by $f_{M,c}$ and $f_{T,c}$, while the available bandwidth -- by $B_{M}$ and $B_{T}$. In what follows, we assume that $B_{T}=\infty$. The rationale behind this assumption is that the bandwidth to cellular systems roughly increases by ten times in each generation and is expected to reach 20+ GHz for 6G THz access. Accounting for limited coverage areas and the rates of forthcoming applications (e.g., 10-40 Gbps for XR/VR, \cite{petrov2020ieee}) the capacity of these systems will be sufficient to maintain thousands of simultaneous sessions. However, the proposed framework can also be modified to account for finite capacities at THz BSs \cite{begishev2021joint}.

\subsection{Blockage, Antenna, and Propagation Models}

\subsubsection{Blockage Model} 

We assume that the LoS between the mmWave/THz BS and the UE can be blocked by users. To this aim, we adopt a dynamic blockage model. According to it, pedestrians, acting as blockers, move by following a random direction mobility model (RDM, \cite{nain2005properties}). The speed and run length are $v_B$ m/s and an $\tau$ m, respectively. The density of pedestrians is $\lambda_B$ units/m$^2$. We model them by cylinders with radius $r_{B}$ and height $h_B$. The height of blockers is assumed to be $h_{B}$, $h_B\leq{}h_U$, where $h_U$ is the constant height of UEs. %In practice, $h_{B}$ is set to be the average height of humans, $1.7$m.

\subsubsection{Antenna Model}

At both BS types we assume planar antenna arrays. By following to \cite{interference1,petrov2017interference}, we use the cone antenna model, where the width of the beam coincides with the radiation pattern's half-power beamwidth (HPBW). The mean gain over HPBW is known to be \cite{constantine2005antenna}
\begin{align}\label{eqn:gain}
G = \frac{1}{\theta_{3db}^+-\theta_{3db}^{-}}\int_{\theta_{3db}^-}^{\theta_{3db}^+} \frac{\sin(N_{(\cdot)}\pi\cos(\theta)/2)}{\sin(\pi\cos(\theta)/2)}d\theta,
\end{align}
where $N_{(\cdot)}$ is the number of antenna elements.

THz BSs and UEs are equipped with arrays having $N_{T,B,V}\times{}N_{T,B,H}$ and $N_{T,U,V}\times{}N_{T,U,H}$ elements, respectively. Similarly, mmWave BS and UEs are equipped with arrays having $N_{M,B,V}\times{}N_{M,B,H}$ and $N_{M,U,V}\times{}N_{M,U,H}$ elements, respectively. The array's HPBW, $\alpha$, is found as $\alpha = 2|\theta_m - \theta_{3db}^{\pm}|$, where $\theta_m=\arccos(-1/\pi)$ is the radiation pattern's maximum, $\theta_{3db}^{\pm}=\arccos[-\pm2.782/(N_{(\cdot)}\pi)]$ are the $\pm$3-dB points. In practice, one may also use approximation $\alpha=102/N_{(\cdot)}$ \cite{constantine2005antenna}. 

\subsubsection{MmWave Propagation Model} 

To represent the mmWave path loss we utilize 3GPP Urban Micro (UMi) Canyon-Street propagation model defined in \cite{standard_16}, i.e.,
\begin{align}\label{eqn:plDb}
\hspace{-2mm}L_{dB}(y)=
\begin{cases}
32.4 + 31.9\log(y) + 20\log{f_{M,c}},\,\text{blocked},\\
32.4 + 21\log(y) + 20\log{f_{M,c}},\,\text{non-blocked},\\
\end{cases}
\end{align}
where $y$ is the UE-BS distance measured in meters while $f_{M,c}$ is the carrier frequency in GHz.

The model in (\ref{eqn:plDb}) can be converted to the linear scale, to the form $A_{M,i}y^{-\zeta_{M,i}}$, where $A_{M,i}$ and $\zeta_{M,i}$ are the some coefficients. Introducing $(A_{M,1},\zeta_{M,1})$ and $(A_{M,2},\zeta_{M,2})$ corresponding to LoS blocked and non-blocked states, we have
\begin{align}\label{eqn:propCoefficient}
A_M=10^{2\log_{10}{f_{M,c}}+3.24},\,\zeta_{M,1}=2.1,\,\zeta_{M,2}=3.19.
\end{align}

Thus, the value of SINR at the mmWave UE can be written as a weighted function of SINRs in blocked and non-blocked states, as follows
\begin{align}\label{eqn:prop_mmWave}
\hspace{-2mm}S(y)=\frac{P_{M}G_{M,B}G_{M,U}[{y}^{-\zeta_{M,1}}[1-p_B(y)]+{y}^{-\zeta_{M,2}}p_B(y)]}{A_{M}(N_0B_M+M_{M,I})},
\end{align}
where $P_{M}$ is mmWave BS emitted power, $G_{M,B}$ and $G_{M,U}$ are the mmWave BS and UE gains, $B_M$ is the bandwidth of mmWave BS, $M_{M,I}$ is the interference margin, $y$ is 3D UE-BS distance (Fig.~ \ref{fig:main}) and $p_B(y)$ is the blockage probability given by \cite{gapeyenko2016analysis}
\begin{align}\label{eqn:blockage}
p_{B}(y)=1-\exp^{-2\lambda_Br_B\left[\sqrt{y^2-(h_{M,B}-h_U)^2}\frac{h_{B}-h_{U}}{h_{M,B}-h_{U}}+r_B\right]},
\end{align}
where $\lambda_B$ is the intensity of blockers, $r_B$ and $h_B$ are the blockers' radius and height, $h_U$ is the height of UE, $h_{M,B}$ is the height of mmWave BS.

\subsubsection{THz Propagation Model}

The principal difference between mmWave and THz propagation models is presence of atmospheric absorption \cite{jornet2011channel}. By utilizing the results of \cite{petrov2017interference}, the absorption loss is defined as 
\begin{equation}\label{eq:l_a}
L_{A}(f,y) = 1/\tau(f_{T,c},y),
\end{equation}
where $\tau(f_{T,c},y)$ is the transmittance of the medium following the Beer-Lambert law, $\tau(f_{T,c},y) \approx e^{-Ky}$, $K$ is the absorption coefficient that can be computed by utilizing the HITRAN database \cite{hitran} as shown in \cite{jornet2011channel}.

Similarly to (\ref{eqn:prop_mmWave}), SINR at the THz UE can be written as
\begin{align}\label{eqn:prop}
\hspace{-2mm}S(y)=\frac{P_{T}G_{T,B}G_{T,U}e^{-Ky}[{y}^{-\zeta_{T,1}}[1-p_B(y)]+{y}^{-\zeta_{T,2}}p_B(y)]}{A_{T}(N_0B_T+M_{T,I})},
\end{align}
where $P_{T}$ is emitted power, $G_{T,B}$ and $G_{T,U}$ are the THz BS and UE gains, $B_T$ is the bandwidth of THz BS, $M_{T,I}$ is the interference margin, $A_T$ is the THz propagation coefficient computed similarly to (\ref{eqn:propCoefficient}), $\zeta_{T,1}$ and $\zeta_{T,2}$ are THz path loss exponents for non-blocked and blocked states, respectively, $p_B(y)$ is the blockage probability provided in (\ref{eqn:blockage}), where the height of mmWave BS is replaced with that of THz BS $h_{T,B}$.

% No interference

%Due to the use of directional antenna radiation patterns utilized at both mmWave and THz BSs the intra-technology interference is assumed to be negligible and modeled by utilizing interference margins, $M_{M,I}$ and $M_{T,I}$. These margins can be calculated using the stochastic geometry models, e.g., \cite{kovalchukov2019evaluating,kovalchukov2019evaluating}.

The assumptions regarding propagation, antenna, blockage and interference models can be further extended to fit the needs of a particular deployment. We will concentrate on crucial factors affecting session-level dynamics by causing potential connection interruptions (micromobility and blockage) or drastic long-term change in the required rate (blockage). Specifically, for a given deployment density, one may estimate it by employing stochastic geometry models \cite{kovalchukov2019evaluating}. Note that the more complex 3GPP multi-path 3D cluster-based model \cite{standard_16} can also be used in the proposed framework as shown in \cite{moltchanov2021performance}. Finally, more comprehensive blockage models capturing blockage by large stationary objects such as buildings can also be used as discussed in \cite{margarita2018multicon}.

\begin{figure*}[!t]
\vspace{-0mm}
\footnotesize
\begin{align}\label{eqn:overallpdf}
\hspace{-3mm}f_{T_A}(t)=\frac{\frac{e^{-\frac{(\log (t)-\mu_x)^2}{2 \sigma_x^2}}}{\sigma_x} \left[2-\text{erfc}\left(\frac{\mu_y-\log (t)}{\sqrt{2} \sigma_y}\right)\right]+\frac{e^{-\frac{(\log (t)-\mu_y)^2}{2 \sigma_y^2}}}{\sigma_y} \left[2-\text{erfc}\left(\frac{\mu_x-\log (t)}{\sqrt{2} \sigma_x}\right)\right]}{ 2   \sqrt{2 \pi } t       \left[1-\frac{1}{2} \text{erfc}\left(\frac{\mu_\phi-\log (t)}{\sqrt{2} \sigma_\phi}\right)+\frac{1}{2} \text{erfc}\left(\frac{\mu_\theta-\log (t)}{\sqrt{2} \sigma_\theta}\right)\right]^{-1}}
+
\frac{\frac{e^{-\frac{(\log (t)-\mu_\phi)^2}{2 \sigma_\phi^2}}}{\sigma_\phi} \left[2-\text{erfc}\left(\frac{\mu_\theta-\log (t)}{\sqrt{2} \sigma_\theta}\right)\right]+\frac{e^{-\frac{(\log (t)-\mu_\theta)^2}{2 \sigma_\theta^2}}}{\sigma_\theta} \left[2-\text{erfc}\left(\frac{\mu_\phi-\log (t)}{\sqrt{2} \sigma_\phi}\right)\right]}{ 2   \sqrt{2 \pi } t       \left[1-\frac{1}{2} \text{erfc}\left(\frac{\mu_x-\log (t)}{\sqrt{2} \sigma_x}\right)+\frac{1}{2} \text{erfc}\left(\frac{\mu_y-\log (t)}{\sqrt{2} \sigma_y}\right)\right]^{-1}}
\end{align}
\normalsize
\hrulefill
\vspace{-4mm}
\end{figure*}

\subsection{Micromobility and Beamalignment Models}

\subsubsection{Micromobility Model}

To represent the micromobility process at UE we utilize the model introduced in \cite{petrov2020capacity}. According to it, the micromobility is modeled as a combination of random displacements processes $x(t)$ and $y(t)$ over $x$- and $y$-axes, together with the random rotation processes over yaw (normal) and pitch (transverse) axes, $\phi(t)$ and $\theta(t)$. By assuming Brownian motion nature of these processes, the authors in \cite{petrov2020capacity} revealed that the probability density function (pdf) of time to outage follows (\ref{eqn:overallpdf}), where $\text{erfc}(\cdot)$ is the complementary error function, $\mu_{(\cdot)}$ and $\sigma_{(\cdot)}$ are the parameters of the corresponding displacement and rotation components that can be estimated from the empirical data provided in \cite{stepanov2021statistical}.

\subsubsection{Beamalignment Model}

% In the considered scenario the decision of how to utilize THz and mmWave interfaces heavily depends on the interplay between beamalignment time and time to outage caused by micromobility. 

% Beamsearch procedures

The ability of applications to tolerate outages caused by micromobility may improve system performance. For this reason, we will explicitly account for the beamalignment time. We consider the hierarchical beamalignment scheme realized via sector scan and in-sector refinement procedures. This approach is utilized in IEEE 802.11ad/ay, where communications entities perform beamalignment separately by forcing the other side to use the omnidirectional mode. The beamalignment time of this approach is $T_B=(N_{T,B,H}N_{T,B,V}+N_{T,U,H}N_{T,U,V})\delta$, where $N_{T,B,H}N_{T,B,V}$ and $N_{T,U,H}N_{T,U,V}$ are the THz BS and UE antenna array configurations and $\delta$ is the array switching time. % Particularly, the BS side sends beacons over all possible array configurations, while the UE measures the received signal strength in the omnidirectional mode. At the second step, these roles are inverted.

% Very small outage time due to micromobility

%The array switching time, $\delta$, is a key parameter that depends on the antenna array implementation and may vary from microseconds to milliseconds. Here, assuming, e.g., $N_{T,B,H}\times{}N_{T,B,V}=64$, $N_{T,U,H}\times{}N_{T,U,V}=4$, $\delta=2$ $\mu$s $T_B\approx{}0.41$ ms which is two orders of magnitude smaller than the typical blockages times \cite{gapeyenko2017temporal}.

\subsection{Associations, Applications, and Multi-connectivity}

\subsubsection{User Association Schemes}

A session associated with UE is assumed to arrive at either mmWave BS or THz BS based on its geometric location and coverage of THz BSs. Particularly, with probability $p_T=K\pi{}r_T^2/\pi{}r_M^2$ session initially arrives to one of the THz BS and with complementary probability $(1-p_T)$ it arrives to mmWave BS, where $K$ is the number of THz BSs, while coverage radius of THz BS, $r_T$, depends on the considered association schemes:
\begin{itemize}
    \item \textit{Outage avoidance, A1.} In this association scheme, only those sessions that do not experience outage in blockage conditions with THz BSs are initially accepted at THz BS. As one may observe, it leads to higher offered traffic load to mmWave BS but may improve service reliability of sessions initially associated with THz BSs, especially, for applications characterized by low micromobility and/or in dense blockage environments.
    \item \textit{Coverage enhancement, A2.} In this association scheme, one attempts to maximize the number of sessions accepted to THz BSs, admitting even those that may experience outage in blockage conditions. This strategy may enhance the coverage of THz BSs and can be suitable for sparse blockers density conditions.
\end{itemize}

\subsubsection{Applications and Multi-connectivity Strategies}

% Applications

We consider rate-greedy applications specifically tailored for 5G/6G systems with mmWave and THz resource-rich wireless access, e.g., AR/VR, 8/16K streaming, holographic telepresence \cite{vannithamby2017towards}. These applications generate non-elastic traffic, i.e., the requested rate cannot be changed during the session. As described below, we also consider applications that may or may not tolerate outage time $T_O=T_B$ caused by micromobility, where $T_B$ is the beamalignment time at THz BS.

% multi-connectivity strategies

The sessions that initially arrive to mmWave BS are assumed to never change their association point while those, initially accepted at THz BSs, have an option to switch to mmWave BS in case of outage caused by either micromobility or outage or both, depending on the considered multi-connectivity strategy. To make conclusions on the optimal UE behavior in joint THz/mmWave radio access network we consider the following multi-connectivity strategies:
\begin{itemize}
    \item \textit{No dynamic multi-connectivity, S1.} For this strategy, we assume that multi-connectivity is not supported at UEs and the initial association point remains unchanged. 
    \item \textit{Blockage avoidance for outage non-sensitive applications, S2.} In this case, we assume that the application may tolerate outages caused by antenna misalignment at THz BS. However, to deal with blockage, UEs support multi-connectivity and the session is rerouted to mmWave BS whenever outage caused by blockage occurs.
    \item \textit{Blockage avoidance for outage sensitive applications, S3.} This strategy is essentially similar to the previous one except we assume that the application cannot tolerate outage caused by micromobility. Thus, whenever micromobility happens, the session is assumed to be lost.
    \item \textit{Fully dynamic multi-connectivity, S4.} In fully dynamic multi-connectivity, we assume that outages caused by both blockage and micromobility lead to rerouting of session associated with THz BS to mmWave BS.
\end{itemize}

In all the considered strategies, where multi-connectivity is supported, the session rerouted to mmWave BS is returned to THz BS, once either the blockage expires or beamalignment is completed. By combining the association schemes and multi-connectivity strategies, one may capture a wide range of potential operational policies in joint THz/mmWave deployments.

\subsection{Traffic, Resources, and Metrics of Interest} \label{sec:regimes}

\subsubsection{Traffic and Resources}

% Resources

We assume that the amount of resources available at mmWave BS is $R$ primary resource blocks (PRB) that depends on the bandwidth $B_{M}$ and the chosen numerology of NR technology \cite{nrmcs}. Contrarily, the amount of resources at THz BS is assumed to be sufficient to handle all the arriving sessions, i.e., virtually unlimited due to large bandwidth $B_T>>B_M$ and small coverage $r_T<<r_M$.

% Traffic

The session arrival process is Poisson with intensity $\lambda_A$ sess./m$^2$. The overall session arrival intensity is $\lambda_A\pi{}r_M^2$. An arriving session is associated with UE and each UE is allowed to have at most one active session. The geometric locations of arriving sessions are assumed to be randomly and uniformly distributed within the coverage area of mmWave BS. Each arriving session is assumed to request the bitrate $C$. The amount of requested resources depends on the location of UE and utilized MCSs and is found in Section \ref{sec:param}.

% Reasons for losses

A session initially arriving to mmWave can be lost due to insufficient amount of resources available. Note that during the service process, a session initially accepted to mmWave BS is also subject to blockage. However, the coverage of mmWave BS is chosen such that these blockage events do not lead to outage. Nevertheless, sessions initially associated with mmWave BS may also be lost during the service as blockage events lead to resource reallocation due to the use of different MCSs. A session initially arriving to THz BS is never lost upon arrival due to abundance of available resources. However, depending on the association scheme and multi-connectivity strategy it might be lost as a result of rerouting to mmWave BS or when no actions taken in case of outage.

\subsubsection{Metrics of Interest}

For the considered system, we are interested in identifying the optimal association scheme and multi-connectivity strategy for different environmental conditions and applications. We will base our conclusions on the following key performance indicators: (i) new session loss probability, i.e., the probability that a newly arriving session is lost at mmWave BS due to the lack of resources, (ii) ongoing session loss probability, i.e., the probability that a session is lost during the service as a result of the lack of resources at mmWave BS while being rerouted or as a result of outage and/or micromobility at THz BS, and, finally, (iii) mmWave BS resource utilization.

\section{Performance Evaluation Framework}\label{sec:perf}

In this section, we formulate our framework. We start with formalization of the general problem. Then, we consider the simplest case when no dynamic multi-connectivity is performed. Further, we address the general case when multi-connectivity is performed in case of outage caused by both micromobility and blockage. Results for other multi-connectivity strategies are finally provided.

\subsection{Model Formalization}

Consider a queuing network with $K+1$ nodes, where nodes $1, 2,\dots,K$ represent THz BSs, while the node $K+1$ models mmWave BS. Node $k$ receives a Poisson arrival flow of sessions with intensity $\lambda_k$, $k=1,2,\dots,K+1$. Service times of the sessions are exponentially distributed with parameter $\mu$. Note that the arrival intensities are obtained as follows
\begin{align}\label{eq:arrivals}
\lambda_{K+1}=(1-p_T)\lambda_A \pi r_M^2,\,\lambda_k= \frac{p_T \lambda_A \pi r_M^2}{K},\,k=1,2,\dots,K,
\end{align}
where $p_T=K\pi{}r_T^2/\pi{}r_M^2$ is the THz association probability.

% Overall system description

Since the amount of resources at THz BSs is assumed to be sufficient to handle arrival traffic load, nodes $1,2,\dots,K$ are modeled as the infinite server queuing systems, and node $K+1$ -- by the loss queuing system with random resource  requirements (ReLS, \cite{naumov2021}) with $N$ servers and $R$ PRBs. Here, servers represent the maximum number of simultaneously supported sessions at a single mmWave BS. Sessions that arrive initially to the node $K+1$ are not redirected to any other node. Each arriving session requires not only a server, but also a random amount of resources according the pmfs $\{ p_{1,r}\}$ and $\{ p_{2,r}\}$, $r\geq 0$ in case of initial arrival and redirection from other nodes, respectively. If the amount of currently available resources is not sufficient to meet the resource requirements upon arrival or redirection of a session, the session is dropped. 

% mmWave BS

Each session that is served on node $K+1$ is associated with a Poisson flow of events with intensity $\nu$ caused by the blockage state changes. Arrival of an event triggers resource reallocation of the session, i.e., the session releases previously occupied resources, generates new amount of required resources and tries to occupy them again. If there are no sufficient free resources at node $K+1$, the session is dropped.

% THz BS

Each session at nodes $1,2,\dots,K$ is associated with two Poisson flows of events with intensities $\nu_B$ and $\nu_M$ that represent the intensity of outages caused by blockage and micromobility, respectively. In what follows, we will refer to them as $b$-type and $m$-type of events. The duration of the outage caused by blockage and micromobility events is approximated by exponential distributions with parameters $\beta_B$ and $\beta_M$, respectively. Depending on the system response to blockage and micromobility events, we consider four multi-connectivity strategies introduced in Section \ref{sec:syst}. Below, we consider user association scheme A2, where the coverage of THz BS is such that users may experience outage in case of blockage. Association scheme A1 is analyzed similarly by recalculating the THz BS association probability $p_T$ and setting the intensity of $b$-type of events $\nu_B$ to zero. Note that the proposed framework can be modified to capture the case of “sub-6 GHz+mmWave” system operation. To produce this scenario, one needs to disable blockage at mmWave system to emulate a microwave BS but still account for blockage at THz BSs that now become mmWave BSs.

\subsection{Fully Dynamic Multi-connectivity Strategy, S4} \label{sec:s4}

\subsubsection{Parametrization}

In this strategy, sessions that originally arrive to nodes $k=1,2,\dots,K$ (THz BSs) switch to node $K+1$ upon arrival of $b$- or $m$-type of events and return back immediately when the blockage or beamalignment is over given that the service time is still not completed. Therefore, there are two arrival flows at node $k$ representing the initial arrivals (primary sessions) and the arrivals returning from node $K+1$ (secondary sessions). According to the memoryless property of the exponential distribution, the residual service times of secondary sessions have the same distribution as the primary sessions. The arrival intensities of the primary and secondary sessions at node $k$  are $\lambda_k$ and $\gamma_k=\gamma_{B,k}+\gamma_{M,k}$, respectively, where $\lambda_k$ is provided in (\ref{eq:arrivals}), while $\gamma_k,\gamma_{B,k},\gamma_{M,k}$ will be defined below. The overall intensity of sessions leaving the system is $\mu+\nu_B+\nu_M$. The mean number of sessions $\bar{N}_k$ at node $k$ is thus
\begin{equation}
 \bar{N}_k=\frac{\lambda_k+\gamma_k}{\mu+\nu_B+\nu_M}.
\end{equation}

\begin{figure}[!t]
\vspace{-0mm}
\centering\hspace{-0mm}
\subfigure[{Original system with events}]{
	\includegraphics[page=1,clip,trim=0cm 24cm 13cm 1cm, width=0.27\textwidth]{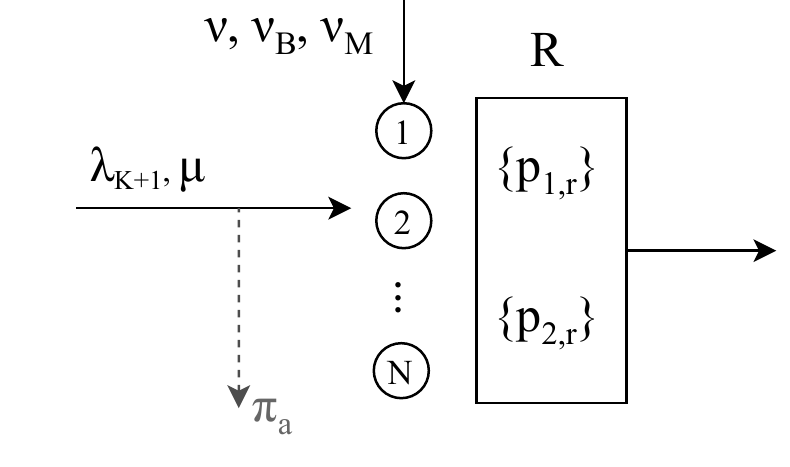}
	\label{fig:queue1}
}\\\vspace{-3mm}
\subfigure[{Equivalent system with additional arrival flows}]{
	\includegraphics[page=1,clip,trim=0cm 22cm 8cm 0.5cm, width=0.4\textwidth]{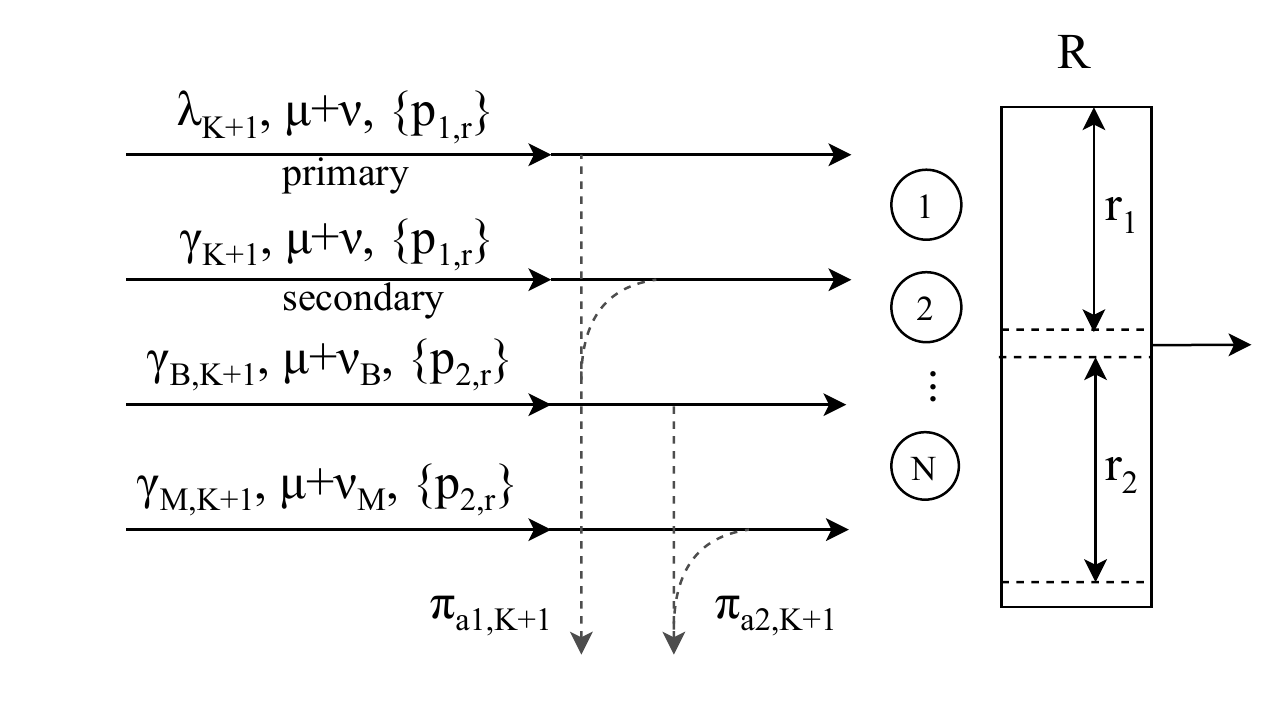}
	\label{fig:queue2}
}
\caption{Original and equivalent ReLS models for node $K$+1.}
\label{fig:queue}
\vspace{-4mm}
\end{figure}

The service process at node $K+1$ (mmWave BS) is described by a loss system with random resource requirements \cite{naumov2016rsmo,naumov2021}. The original service model and the proposed queuing formalization are illustrated in Fig. \ref{fig:queue}. In the considered multi-connectivity strategy, node $K+1$ serves the sessions that originally arrive to it, as well as the sessions that are rerouted from nodes $k=1,2,\dots,K$ during outage periods. Only those sessions that originally arrive to the node $K+1$ suffer from blockage at this node and, thus, similarly to the previously considered strategy, the resource reallocation upon arrival of $b$-type of events is modeled by additional arrival flow of secondary session. Overall, we define the following arrival flows: (i) initial arrivals (primary sessions) with intensity $\lambda_{K+1}$, (ii) arrivals of secondary sessions with intensity $\gamma_{K+1}=\bar{N}_{K+1} \nu $, and (iii) arrivals of rerouted sessions from nodes $k=1,2,\dots,K$ as a result of outage caused by either blockage with intensity $\gamma_{B,K+1}$ or micromobility with intensity $\gamma_{M,K+1}$. The secondary session arrival intensity equals to the total intensity of $b$-type event arrivals, that is,
\begin{align} \label{eq:gammak}
\gamma_{K+1}=\bar{N}_{1,K+1} \nu 
\end{align}

The intensities $\gamma_{B,K+1}$ and $\gamma_{M,K+1}$ are obtained as the sum of all rerouting intensities from nodes $k=1,2,...,K$, i.e.,
\begin{align}\label{eq:gammabk}
\gamma_{B,K+1}=\sum_{k=1}^K \bar{N}_k \nu_B,\,
\gamma_{M,K+1}=\sum_{k=1}^K \bar{N}_k \nu_M.
\end{align}

The service intensity of primary and secondary sessions is $\mu+\nu$, while the service intensities of the rerouted sessions are $\mu+\nu_B$ and $\mu+\nu_M$. The resource requirements of the primary and secondary sessions obey the pmf $\{ p_{1,r}\}$, $r\geq 0$, while those of rerouted sessions follow the pmf $\{ p_{2,r}\}$, $r\geq 0$. Since the resource requirements of primary and secondary sessions are equal, they can be aggregated in a single arrival flow with the offered traffic load
\begin{align}
\rho_1=\frac{\lambda_{K+1}+\gamma_{K+1}}{\mu+\nu}.
\end{align}

The rerouted sessions are characterized by the same resource requirements distribution, so they may also be aggregated in one flow with the offered load
\begin{align}
\rho_2=\frac{\gamma_{B,K+1}}{\mu+\beta_B}+\frac{\gamma_{M,K+1}}{\mu+\beta_M}.
\end{align}

\subsubsection{Solution and Metrics}

The following theorem holds.

\vspace{-2mm}

\begin{thm} 
The stationary probabilities $q_{n_1,n_2}(r_1,r_2)$ that there are $n_1$ primary or secondary sessions occupying $r_1$ resources and $n_2$ rerouted sessions that totally occupy $r_2$ resources have the following form \cite{naumov2016rsmo}
\begin{align} \label{eq:q}
q_{n_1,n_2}(r_1,r_2)&=q_0  \frac{\rho_1^{n_1}}{n_1!}\frac{\rho_2^{n_2}}{n_2!} p_{1,r_1}^{(n_1)} p_{2,r_2}^{(n_2)}, \\
&0 \leq n_1+n_2 \leq N, 0 \leq r_1+r_2 \leq R, \nonumber
\end{align}
where the probability $q_0$ is computed as
\begin{align} \label{eq:q0}
q_0=\left( \sum_{0 \leq n_1+n_2 \leq N} \frac{\rho_1^{n_1}}{n_1!}\frac{\rho_2^{n_2}}{n_2!} \sum_{0 \leq r_1+r_2 \leq R}  p_{1,r_1}^{(n_1)} p_{2,r_2}^{(n_2)} \right)^{-1}.
\end{align}
\end{thm}

\begin{proof}
The proof is provided in \cite{naumov2016rsmo}.
\end{proof}

Having obtained the stationary distribution, we can evaluate the considered performance metrics. Particularly, the mean number of sessions of each aggregated flow is given by
\begin{align} \label{eq:barn12}
\bar{N}_{i,K+1}= \sum_{0 \leq n_1+n_2 \leq N} \sum_{0 \leq r_1+r_2 \leq R} n_i q_{n_1,n_2}(r_1,r_2),\quad i=1,2.
\end{align}

The loss probabilities of sessions initially associated with $K+1$ node (mmWave BS) and of rerouted sessions from node $k$ (THz BSs) are then provided by
\begin{align} \label{eq:pia1}
\pi_{a1,K+1}= 1 - \hspace{-3mm} \sum_{0 \leq n_1+n_2 \leq N-1} \sum_{0 \leq r_1+r_2 \leq R} \hspace{-3mm} q_{n_1,n_2}(r_1,r_2) \hspace{-2mm} \sum_{j=0}^{R-r_1-r_2} p_{1,j}, \nonumber\\
\pi_{a2,K+1}= 1 - \hspace{-3mm} \sum_{0 \leq n_1+n_2 \leq N-1} \sum_{0 \leq r_1+r_2 \leq R} \hspace{-3mm} q_{n_1,n_2}(r_1,r_2) \hspace{-2mm} \sum_{j=0}^{R-r_1-r_2} p_{2,j}.
\end{align}

Denote by $\bar{N}_{2B,K+1}$ the mean number of sessions that are rerouted to node $K+1$ due to blockage, and by $\bar{N}_{2M,K+1}$ -- the mean number of sessions that are rerouted to node $K+1$ due to micromobility. Then, the arrival intensity of secondary sessions at node $k$ (i.e., the intensity of sessions returning back to their original node $k$) is given by
\begin{align} 
\gamma_k=\frac{\bar{N}_k \nu_B}{\gamma_{B,K+1}} \bar{N}_{2B,K+1} \beta_B +\frac{\bar{N}_k \nu_M}{\gamma_{M,K+1}} \bar{N}_{2M,K+1} \beta_M,
\end{align}
where $\bar{N}_k \nu_B/\gamma_{B,K+1}$ and $\bar{N}_k \nu_M/\gamma_{M,K+1}$ are the fractions of sessions that are rerouted from node $k$ to node $K+1$, while $\bar{N}_{2B,K+1} \beta_B$ and $\bar{N}_{2M,K+1} \beta_M$ are the total intensities of sessions returning back to their original node from node $K+1$. The mean number of sessions can be estimated by equating session acceptance and session departure rates, i.e.,
\begin{align}
\gamma_{B,K+1}(1-\pi_{a2,K+1})&=\bar{N}_{2B,K+1} (\mu+\beta_B), \\
\gamma_{M,K+1}(1-\pi_{a2,K+1})&=\bar{N}_{2M,K+1} (\mu+\beta_M),
\end{align}
which leads to (see Appendix)
\begin{align} \label{eq:gamma_k}
\gamma_k=\frac{\bar{N}_k \nu_B (1-\pi_{a2,K+1}) \beta_B}{\mu+\beta_B}  + \frac{\bar{N}_k \nu_M(1-\pi_{a2,K+1}) \beta_M}{\mu+\beta_M}.
\end{align}

The probability that a session initially arriving at node $K+1$, is lost upon arrival is given by the loss probability of the primary sessions, i.e.,
\begin{align}
\pi_N=\pi_{a1,K+1}.
\end{align}

The probability that a session initially arriving at node $K+1$, is lost during the service is evaluated similarly to \eqref{eq:pisk_baseline}, that is,
\begin{align}
\hspace{-1mm} \pi_{s,K+1}=\lim_{t \to \infty} \frac{\bar{N}_{1,K+1} \nu \pi_N t}{\lambda_{K+1}(1-\pi_N) t} = \frac{\bar{N}_{1,K+1} \nu \pi_N}{\lambda_{K+1}(1-\pi_N)}.
\end{align}

The mean number of occupied resources at node $K+1$ is
\begin{align}
\bar{R}=\sum_{1 \leq n_1+n_2 \leq N} (n_1+n_2) \sum_{0 \leq r_1+r_2 \leq R} q_{n_1,n_2}(r_1,r_2).
\end{align}

The sessions that originally arrive at nodes $k=1,2,...,K$ can be lost only during rerouting to node $K+1$. So, their loss probability $\pi_{s,k}$ is equal to the ratio of lost sessions to the accepted sessions as time $t\rightarrow\infty$, i.e.,
\begin{align}
\pi_{s,k} \hspace{-1mm}=\hspace{-1mm} \lim_{t \to \infty} \frac{\bar{N}_k (\nu_B + \nu_M) \pi_{a2,K+1} t}{\lambda_k t} =\frac{\bar{N}_k (\nu_B + \nu_M) \pi_{a2,K+1}}{\lambda_k}.
\end{align}

Finally, by averaging over all the nodes, the total ongoing session loss probability $\pi_O$ is obtained according to \eqref{eq:pio}.

\subsection{No Dynamic Multi-connectivity Strategy, S1}

\subsubsection{Parameterization}

This strategy is a degenerate case of S3, as the service processes at all nodes are independent of each other. Then, the probabilities $b_k$ and $m_k$ of session loss upon arrival of a $b$- or $m$-type of events at nodes $k=1,2,...,K$ are equal to the probability that the outage time exceeds the outage tolerance time, $T_O$,
\begin{align} \label{eq:b_k}
b_k=e^{-\beta_B T_O}, \, m_k=e^{-\beta_M T_O}, \, k=1,2,\dots,K.
\end{align}

The total intensity of sessions leaving the nodes $k=1,2,\dots,K$ is $\mu+\nu_B b_k +\nu_M m_k$, where $\mu$ is the intensity of service completions, $\nu_B b_k$ and $\nu_M m_k$ are the service interruption intensities caused by blockage and micromobility, respectively. The probability $\pi_{s,k}$ of session loss during the service is
\begin{align} \label{eq:pi_sk}
\pi_{s,k}=\frac{\nu_B b_k +\nu_M m_k}{\mu+\nu_B b_k +\nu_M m_k}, \quad k=1,2,...,K.
\end{align}

The service process at node $K+1$ can be described by a loss system with random resource requirements \cite{naumov2016rsmo}, where sessions occupy a server and a random amount of resources. Recall, that sessions at the node $K+1$ are associated with $b$-type of events only. Upon arrival of an event, a session releases the occupied resources, generates new resource requirements according to the same pmf $\{ p_{1,r}\}$, $r\geq 0$ and tries to re-enter the system. If the node $K+1$ has sufficient amount of unoccupied resources to meet the new resource requirements, it is accepted for further service. Otherwise, it is lost during the service.

% By utilizing the approach in \cite{ecms2019}, we approximate the initial ReLS with the service interruption events by an equivalent ReLS without these events but with additional arrival flow of secondary sessions that try to re-enter the system after event arrivals. In this case, the intensity of sessions leaving the system is $\mu+\nu$ and arrival intensity of secondary sessions is $\bar{N}_{K+1} \nu$, where $\bar{N}_{K+1}$ is the mean number of sessions at the node $K+1$.

\subsubsection{Solution and Metrics} 
The following theorem holds true.

\vspace{-2mm}

\begin{thm}
The mean number of sessions at the node $K+1$ (mmWave BS) obeys
\begin{align} \label{eq:N}
\bar{N}_{K+1}=q_0\sum_{n=0}^N n \frac{(\lambda_{K+1}+\bar{N}_{K+1} \nu)^n}{(\mu+\nu)^n n!} \sum_{r=0}^R p_{1,r}^{(n)}, 
\end{align}
where the probability $q_0$ is provided by
\begin{align} 
q_0=\left( 1+ \sum_{n=1}^N \frac{(\lambda_{K+1}+\bar{N}_{K+1} \nu)^n}{(\mu+\nu)^n n!} \sum_{r=0}^R p_{1,r}^{(n)} \right)^{-1},
\end{align}
$p_{1,r}^{(n)}$ is the probability that $n$ sessions occupy $r$ resources that can be obtained from the pmf $\{ p_{1,r}\}$, $r\geq 0$ using convolution. 
\end{thm}

\vspace{-2mm}

\begin{proof}
The proof is provided in \cite{naumov2016rsmo}.
\end{proof}

\vspace{-2mm}

%In practice, $\bar{N}_{K+1}$ is evaluated recursively. At the first iteration, $\bar{N}_{K+1}=0$ is inserted to the right-hand-side of \eqref{eq:N} to obtain new value of $\bar{N}_{K+1}$. At further iterations, the new value of the mean number of sessions replaces the previous one in \eqref{eq:N} until the desired accuracy is obtained. It has been shown that the algorithm converges \cite{ecms2019}.

We are now in position to proceed with the metrics of interest. Since the resource requirements of secondary sessions are characterized by the same pmf as the primary ones, the new session loss probability $\pi_N$ is given by
\begin{align}\label{eq:pia}
\pi_N=1-q_0\sum_{n=0}^{N-1} \frac{(\lambda_{K+1}+\bar{N}_{K+1} \nu)^n}{(\mu+\nu)^n n!} \sum_{r=0}^R p_{1,r}^{(n+1)},
\end{align}
while the ongoing session loss probability at node $K+1$ is 
\begin{align} \label{eq:pisk_baseline}
\pi_{s,K+1}=\lim_ {t \to \infty}\frac{\bar{N}_{K+1} \nu \pi_N t}{\lambda_{K+1}(1-\pi_N) t} = \frac{\bar{N}_{K+1} \nu \pi_N}{\lambda_{K+1} (1-\pi_N)},
\end{align}
where $\bar{N}_{K+1} \nu \pi_N t$ is the number of the secondary sessions lost during time $t$ and $\lambda(1-\pi_N) t$ is the number of accepted sessions during time $t$. By averaging over mmWave and THz sessions, one obtains the ongoing session loss probability as
\begin{align} \label{eq:pio}
    \pi_O & =\frac{\lambda_{K+1}(1-\pi_N)}{\lambda_{K+1}(1-\pi_N)+\sum_{k=1}^K \lambda_k} \pi_O(mmW)+ \nonumber \\
    &+\frac{\sum_{k=1}^K \lambda_k}{\lambda_{K+1}(1-\pi_N) +\sum_{k=1}^K \lambda_k} \pi_O(THz),
\end{align}
where $\pi_O(mmW)$ is the ongoing mmWave session loss probability, which is given by~\eqref{eq:pisk_baseline}, and $\pi_O(THz)$ is the ongoing THz session loss probability, which is derived by averaging over all $K$ THz nodes, i.e.,
\begin{equation}
    \pi_O(THz)=\frac{\sum_{k=1}^K \lambda_k \pi_{s,k}}{\sum_{k=1}^K \lambda_k}
\end{equation}

Finally, the mean number of occupied resources $\bar{R}$ at the node $K+1$ (mmWave BS) is
\begin{align}\label{eq:R}
\bar{R}= q_0\sum_{n=0}^N \frac{(\lambda_{K+1}+\bar{N}_{K+1} \nu)^n}{(\mu+\nu)^n n!} \sum_{r=0}^R r p_{1,r}^{(n)}.
\end{align}

Note that the sums in \eqref{eq:N}, \eqref{eq:pia}, and \eqref{eq:R} are evaluated using the convolution algorithm \cite{sopin2018algorithm}. The user association scheme A1 is analyzed by setting $\nu_B=0$ and recalculating $r_T$.

\subsection{Blockage Avoidance Strategies, S2/S3}

In the blockage avoidance strategies, the system behavior is similar to the fully dynamic multi-connectivity strategy. The difference is that the sessions at the nodes $k=1,2,\dots,K$ are not rerouted to the $K+1$ node upon $m$-type event arrivals implying that $\gamma_{M,K+1}=0$. Besides, the arrival intensity $\gamma_k$ of secondary sessions at node $k$ in \eqref{eq:gamma_k} takes the form
\begin{align} \label{eq:gamma_k_s23}
\gamma_k=(\bar{N}_k \nu_B (1-\pi_{a2,K+1}) \beta_B)/(\mu+\beta_B),
\end{align}
where the mean number of sessions at the nodes $k=1,2,\dots,K$ is evaluated as follows
\begin{align}
\bar{N}_k=(\lambda_k+\gamma_{k})/(\mu+\nu_B+\nu_M m_k),
\end{align}
where $m_k$ is the probability that the beamalignment time exceeds the outage tolerance time $T_O$. Note that for outage non-sensitive sessions (strategy S2) we have $m_k=0$, while $m_k=1$ otherwise (strategy S3). 

For the metrics of interest, the only difference from Section \ref{sec:s4} is the loss probability $\pi_{s,k}$ during the service of those sessions that initially arrive to the nodes $k=1,2,\dots,K$. For the considered strategy, it takes the following form
\begin{align}
\pi_{s,k}=\bar{N}_k (\nu_B \pi_{a2,K+1} + \nu_M m_k) /\lambda_k.
\end{align}

All other expressions and metrics remain unchanged.

\section{Framework Parameterization}\label{sec:param}

In this section, the developed performance evaluation framework is parameterized. The required parameters include: (i) the coverage radii of mmWave and THz, $r_M$ and $r_T$, (ii) the pmfs of the amount of requested resources by a session at mmWave BS, $\{ p_{1,r}\}$ and $\{ p_{2,r}\}$, $r\geq 0$, (iii) the intensity of UE state changes between LoS blocked and non-blocked states at mmWave and THz BS, $\nu$ and $\nu_B$, and (iv) the intensity of state changes at THz BS caused by micromobility $\nu_M$.

\subsection{Effective Coverage Radii}

\subsubsection{MmWave BS Coverage}

% mmWave BS

The coverage of mmWave BSs, $r_M$, is determined by the deployment density in a given environment and is upper bounded by the maximal feasible coverage radius. Since THz systems are expected to be deployed at the mature phase of 5G mmWave NR deployments, we assume a well-provisioned deployment of mmWave BSs. In these conditions $r_M$ can be determined such that no more than $p_{M,O}<<1$, fraction of cell edge UEs are in outage conditions.

% How the procedure proceeds

Consider the SINR outage threshold $S_{M,\min}$ representing the minimum SINR value for NR MCS schemes \cite{nrmcs}. Using the propagation model in (\ref{eqn:prop_mmWave}), we can write
\begin{align}\label{eqn:smin}
S_{M,\min}=\frac{P_M G_{M,B} G_{M,U}}{A_{M}(N_0B_M+M_{M,I}) }(r_M^2+[h_{M,B}-h_U]^2)^{-\zeta_{M,2}/2},
\end{align}
where $\zeta_{M,2}$ is the path loss exponent in LoS blocked state, $h_{M,B}$ and $h_U$ are the heights of mmWave BS and UE, $P_M$ is the mmWave BS transmit power, $G_{M,B}$ and $G_{M,U}$ are the mmWave BS transmit and the UE receive antenna gains, $A_{M}$ is the propagation coefficient, $M_{M,I}$ is the interference margin, $N_0$ is the thermal noise and $B_M$ is the mmWave BS bandwidth.
	
Solving (\ref{eqn:smin}) with respect to $r_M$ yields
\begin{align}
r_M=
\sqrt{\left(\frac{P_MG_{M,B}G_{M,U}}{A_{M}(N_0B_M+M_{M,I})S_{M,\min}}
\right)^{2/\zeta_{M,2}}\hspace{-6mm}-(h_{M,B} - h_U)^2},
\end{align}
where $M_{M,2}$ is the shadow fading margin in the blocked state, which is computed as follows
\begin{align}
M_{M,2}=\sqrt{2}\sigma_{M,2}\text{erfc}^{-1}(2[1-p_{M,O}]),
\end{align}
where $\text{erfc}^{-1}(\cdot)$ is the inverse complementary error function, $p_{M,O}$ is the cell-edge outage probability, and $\sigma_{M,2}$ is the standard deviation of the shadow fading distribution in the LoS blocked state, which is provided in \cite{standard_16}.

\subsubsection{THz BS Coverage}

For a given $r_M$, the coverage radius of THz BS, $r_T$, heavily affects THz BS association probability, $p_T=K\pi{}r_T^2/\pi{}r_M^2$. Having higher $r_T$ leads to better offloading gains to bandwidth-rich THz BSs but may also result in higher session losses. For user association scheme A1, where only those sessions that do not experience outage in blockage conditions are initially accepted to THz BSs, the coverage radius $r_T$ is computed similarly to mmWave BS above. For coverage enhancement association scheme A2, one needs to replace the propagation model with the model corresponding to LoS non-blocked conditions in the calculations of the THz BS coverage radius $r_T$.

\subsection{Resource Request Characterization}

%The coverage of mmWave NR and THz BSs affects the amount of resources requested by a session. Thus, accounting for the propagation model, we proceed with deriving the probability mass function (pmf) of the number of requested resources. Particularly, for mmWave BS we determine the sought pmf by first establishing the pmf of the number of requested resources in the LoS non-blocked and blocked states, and then weighing them with the corresponding probabilities. Recalling that the SINR values in the LoS non-blocked and blocked conditions differ only by a constant factor, we provide a detailed derivation of the pmf for the LoS non-blocked conditions only. 

% Distribution of the distance

\subsubsection{Sessions Initially Arriving to mmWave BS}

\begin{figure}[b!]
\vspace{-4mm}
\centering
\includegraphics[width=.5\columnwidth]{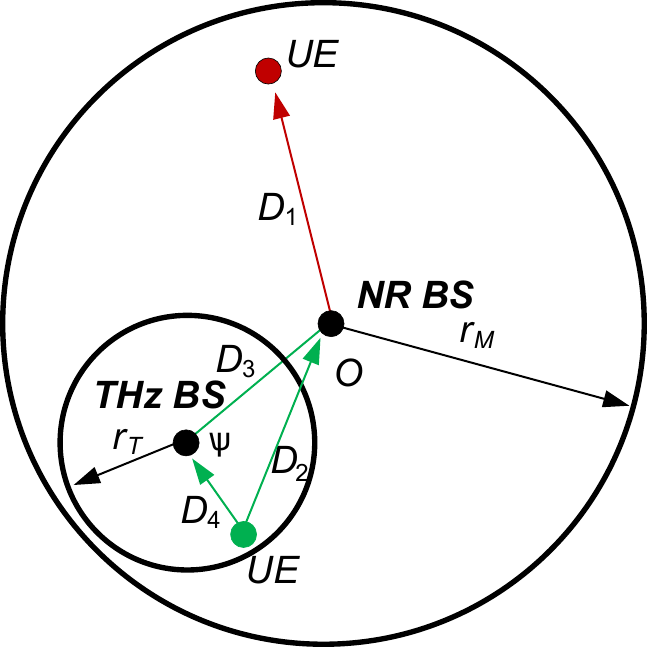}
\caption{Distances in the considered deployment.}
\label{fig:distances}
\vspace{-0mm}
\end{figure}

Consider first sessions that are initially associated with mmWave BS, see Fig. \ref{fig:distances}. Recall that geometric locations of these sessions are uniformly distributed in the coverage area of mmWave BS. Hence, two-dimensional distance to the mmWave BS can be approximated by  $f_{D_1}(x)=2x/r_M^2$, $0<x<r_M$. Therefore, the pdf of the 3D distance, $Y_1$, is provided by
\begin{align}
f_{Y_1}(y)=2y/r_M^2,\,y\in(|h_{M,B}-h_U|,Q),
\end{align}
where $Q=\sqrt{r_M^2+(h_{M,B}-h_U)^2}$ leading to the CDF $F_{Y_1}(y)$ in the following form of
\begin{align}\label{FDy}
F_{Y_1}(y)=\frac{y^2-(h_{M,B}-h_U)^2}{r_M^2},\,y\in[|h_{M,B}-h_U |,Q].
\end{align}

% SNR distribution in non-blockage conditions

Observe that SINR is a decreasing function of $y$. Thus, SINR CDF can be written by utilizing the distance $Y_1$. We have
\begin{align}\label{FSnB}
\hspace{-2mm}F_{S_{nB}}(s)&=Pr\left\{C_{M,1} y^{-\zeta_{M,1}}<s\right\}=1-F_{Y_1}\left(\sqrt[\zeta_{M,1}]{C_{M,1}/s}\right),
\end{align}
where $C_{M,1}=P_MG_{M,B}G_{M,U}/A_{M}(N_0B_M+M_{M,I})$.

Now, the SINR CDF in LoS non-blocked state is given by
\begin{align}\label{F_S_Bn(s)}
\hspace{-4mm}F_{S_{nB}}(s) = \frac{Q^2 - \left(\frac{C_{M,1}}{s}\right)^{\frac{2}{\zeta_{M,1}}}}{r_M^2},\,\frac{C_{M,1}}{Q^{\zeta_{M,1}}} \leq s < \frac{C_{M,1}}{(h_{M,B}-h_U)^{\zeta_{M,1}}}.
\end{align}

% To determine the overall SINR CDF, we now need the probability of blockage provided in (\ref{eqn:blockage}).

% Weighting the getting resources

The CDF $F_{S_{B}}(s)$ of the random variable (RV) $S_{B}$ denoting the SINR in the LoS blocked state is obtained similarly. The final SINR CDF is obtained by weighting individual branches with the blockage probability $p_B(y)$ provided in (\ref{eqn:blockage}). Finally, we define  $S_j, j=1,2,\dots,J$, to be SINR boundaries for MCS mapping \cite{nrmcs} and also let $\epsilon_j$ be the probability that the session is associated with MCS $j$ and requires $r_j$ PRBs. By utilizing the SINR CDF $F_{S}(s)$, we have
\begin{align}
\epsilon_j=F_{S}(S_{j+1})-F_{S}(S_j),\, j=1,2,\dots,J,
\end{align}
and the probability $\epsilon_j$ that a session requests $r_j$ PRBs can now be used to obtain the resource requirements pmf $\{ p_{1,r}\}$ $r\geq{}0$.

\subsubsection{Sessions Initially Arriving to THz BS}

% Resources by UE associated with THz BS

Consider now UE that is initially associated with THz UE and determine its resource requirements at mmWave BS. The only principal difference compared to the abovementioned analysis is that the pdf of 2D distance from UE to mmWave BS, $f_{D_2}(x)$, is different from $f_{D_1}(x)$. It can be found by applying the cosine theorem to the triangle organized by sides $D_2$, $D_3$, and $D_4$, as illustrated in Fig. \ref{fig:distances}, that is,
\begin{align}\label{eqn:d2}
D_2=\sqrt{D_3^2+D_4^2-2D_3D_4\cos\psi},
\end{align}
where pdf of components are given by
\begin{align}
f_{D_3}=2x/r_M^2,\, f_{D_4}=2x/r_T^2,\, f_{\psi}(x)=1/2\pi.
\end{align}

The pdf $f_{D_2}(x)$, corresponding to
(\ref{eqn:d2}) can be obtained by using the non-linear transformation of RVs, see, e.g., \cite{ross}. Once $f_{D_2}(x)$ is obtained the resource request pmf at mmWave BS of UEs that are initially associated with THz BSs $\{p_{2,r}\}$ $r\geq{}0$, can be determined similarly to that of UEs initially associated with mmWave BS.

\subsection{Intensities of UE State Changes}

Recall, that sessions that are initially associated with THz BSs are allowed to utilize multi-connectivity functionality by switching to mmWave BS in case of outages. The latter may occur as a result of two phenomena: (i) blockage and (ii) micromobility. Let $\nu_B$ and $\nu_M$ be the corresponding intensities of these events. Note that those sessions that are initially associated with mmWave BS are also subject to blockage events with intensity $\nu$.

\subsubsection{Blockage Intensity at THz/mmWave BSs}

% Blockage process

We first need to convert the spatial density of blockers, $\lambda_B$, into temporal intensity of blockers entering the blockage zone $\alpha$. By utilizing the results in \cite{gapeyenko2016analysis} we first determine the mean perimeter of the LoS blockage zone as follows
\begin{align}
P_L(x)=4r_B+2x(h_{B}-h_{U})(h_{T,B}-h_{U}),
\end{align}
where $x$ is 2D distance between BS and UE.

Now, consider the unit area around the LoS blockage zone. A blocker located in this zone and moving according to the RDM model with speed $v_B=1$ m/s enters the LoS blockage area in approximately $2/5$ of cases in a unit time. Thus, the temporal intensity of blockers entering the LoS blockage zone associated with UE located at 2D distance $x$ from THz BS is
\begin{align}
\alpha(x)=2\lambda_Bv_B[4r_B+2x(h_{B}-h_{U})(h_{T,B}-h_{U})]/5.
\end{align}

% Mean values of outage non-outage

Let further $\Psi_B(x)$ and $\Phi_B(x)$ be the random blocked and non-blocked periods. It has been shown in \cite{gapeyenko2017temporal} that $\Phi_B(x)$ following exponential distribution with parameter $\alpha(x)$, i.e., $E[\Phi_B(x)]=1/\alpha(x)$. The non-blocked period duration coincides with the busy period in M/G/$\infty$ queuing system with arrival intensity $\alpha(x)$ and service times corresponding to the time a single blocker spends in the LoS blockage zone \cite{gapeyenko2017temporal}. Recalling that the length of the LoS blockage zone is much greater than its width, the latter can be approximated by $2r_B/v_B$. By further utilizing M/M/$\infty$ approximation for M/G/$\infty$ we can write down closed-form approximation for the mean LoS blocked period in the following form \cite{cohen2012single}
\begin{align}
E[\Psi(x)]=(e^{-\alpha(x)v_B/2r_B}-1)/\alpha(x).
\end{align}

% Intensity

In the outage avoidance association scheme A1 no UEs experience outage conditions as a result of blockage at THz BS. For, coverage enhancement scheme A2, the intensity of UE state changes leading to outage can be found as
\begin{align}
\nu_B=\int_{r_{T,1}}^{r_{T,2}}(\alpha^{-1}(x)+[\alpha^{-1}(x)e^{-\alpha(x)v_B/2r_B}-1])^{-1}dx,
\end{align}
where $r_{T,1}$ and $r_{T,2}$ are THz BS coverage radii corresponding to A1 and A2 association schemes. Note that intensity of the blockage process at mmWave BSs can be found similarly except for the lower integration limit that has to be set of the minimum separation distance between mmWave BS and UE.

% micromobility process

\subsubsection{Micromobility Intensity}

The intensity of UE outages at THz BS caused by micromobility can be found by utilizing the results of \cite{petrov2020capacity}. For on-demand and periodic beamalignment schemes, the reported fractions of time in outage are
\begin{align}
&p_{O,1}=\int_{0}^{\infty}\frac{T_B}{t+T_\text{B}}f_{T_A}(t)dt,\nonumber\\
&p_{O,2}=\frac{T_B[1-F_{T_A}(T_{U})]}{T_U+T_B}+\int_{0}^{T_U}\frac{T_U+T_B-t}{T_U+T_B}f_{T_A}(t)dt,
\end{align}
correspondingly, where $T_B$ is the beamalignment time, $T_U$ is the regular beamalignment interval, $f_{T_A}(t)$ is the time to outage provided in (\ref{eqn:overallpdf}), while $f_{T_A}(T_U)$ is given by
\begin{align}\label{eqn:q}
F_{T_A}(T_{U})=Pr\{T_A<T_U\}=\int_{0}^{T_U}f_{T_A}(t)dt,
\end{align}
and represents the probability that the time to outage is smaller than the beamalignment interval.

For on-demand beamalignment, the intensity of UE state changes is thus $\nu_M=p_{O,1}/T_B$. For periodic beamalignment, however, the outage time does not necessarily coincide with the beamalignment time as the non-outage time $T_A$ might be smaller than $T_U$. The mean outage time in this case is
\begin{align}\label{eqn:t02}
T_{\text{O},2}=
\begin{cases}
T_B/(T_U+T_B),&T_A\geq{}T_U,\\
(T_U+T_B-T_A)/(T_U+T_B),&T_A<T_U,\\
\end{cases}
\end{align}
leading to the following intensity of UE state changes
\begin{align}
\nu_M=\frac{\frac{T_B[1-F_{T_A}(T_{U})]}{T_U+T_B}+\int_{0}^{T_U}\frac{T_U+T_B-t}{T_U+T_B}f_{T_A}(t)dt}{\int_{0}^{T_U}(T_U+T_B-t)f_{T_A}(t)dt+[1-F_{T_A}(T_{U})]T_B}.
\end{align}

	\begin{table} [!t]
	\vspace{-0mm}
	\centering
	\caption{The default system parameters.}
	\label{table:numParam}
	\begin{tabular}{l l l}
		\hline 
		\textbf{Notation}& \textbf{Description}& \textbf{Values}\\ 
		\hline 
		\hline
		$f_{M,c}$ & mmWave carrier frequency & 28 GHz\\
		\hline
		$f_{T,c}$ & THz carrier frequency & 300 GHz\\
		\hline
		$B_{M}$ & mmWave BS bandwidth & 400 MHz\\
		\hline
		$C$ & required session rate & 10 Mbps\\
		\hline
		$\mu^{-1}$ & mean session service time & 10 s\\ 
		\hline
		$K$ & number of THz BSs & 3-8\\
		\hline
		$\lambda_A$& session arrival intensity & $10^{-4}$ sess./s/m$^2$\\
		\hline
		$\lambda_B$& density of blockers & 0.1 bl./m$^2$\\  
		\hline
		$r_B$ & blocker radius & 0.4 m\\
		\hline
		$h_{T,B}$ & THz BS height & 10 m\\ %10 m
		\hline 
		$h_{M,B}$ & mmWave BS height  & 10 m\\
		\hline
		$h_U$ &  UE height& 1.7 m\\
		\hline
		$v_B$ & UE speed & 1 m/s\\ 
		\hline
		$P_M,P_T$ & mmWave/THz BS emitted power & 2 W\\
		\hline
		$\zeta_{M,1},\zeta_{T,1}$ &  path loss exponents in non-bl. state & 2.1\\
		\hline
		$\zeta_{M,2},\zeta_{T,2}$ &  path loss exponents in bl. state & 2.1\\
		\hline
		$M_{M,I},M_{T,I}$ & mmWave/THz interference margins & 3 dBi\\
		\hline
		$\sigma_{M,2}$ & STD of shadow fading in bl. state & 7.2 dBi\\
		\hline
		$N_{T,B,V},N_{T,B,H}$ & THz BS antenna configuration & 16$\times$4 \\
		\hline
		$N_{T,U,V},N_{T,U,H}$ & THz BS antenna configuration & 4$\times$4 \\
		\hline
		$N_{M,B,V},N_{M,B,H}$ & mmWave BS antenna configuration & 8$\times$4 \\
		\hline
		$N_{M,U,V},N_{M,U,H}$ & mmWave UE antenna configuration & 4$\times$4 \\
		\hline
		$N_0$ & thermal noise power & -84 dBi \\
		\hline 
		$\Delta{x},\Delta{y}$ & mean displacement over $0x$ and $0y$ & 0.03 m/s \\
		\hline 
		$\Delta\phi, \Delta\theta$ & mean yaw/pitch displacement & 0.1$^{\circ}$/s \\
		\hline
		$\beta_B$& recovery intensity for blockage & 1.25 events/s\\
		\hline
		$\beta_M$ & recovery intensity for micromob. & 1800 events/s\\
		\hline
		$T_B$ & beamalignment time & 1/$\beta_M$ s \\
		\hline
		$T_O$ & application outage tolerance time & $=T_B$ s\\
		\hline
%		\textcolor{blue}{$T_U$} &  \textcolor{blue}{periodic beamalignment interval}& ..., s\\
%		\hline
%		\textcolor{red}{$\beta_B$} & \textcolor{red}{intensity of connection recovery after blockage} & 1.25 1/s\\
%		\hline
		$p_{M,O},p_{T,O}$ & mmWave/THz cell edge outage & $0.05$\\
		\hline
        $J$ & number of MCSs & 15 \\
\hline
	\end{tabular} 
	\vspace{-4mm}
    \end{table}

%Finally, the intensity of UE states changes is $\nu=\nu_B+\nu_M$.

\begin{table} [!b]
	\vspace{-4mm}
	\centering
	\caption{Service radius of THz and mmWave BS}
	\label{table:radii}
	\begin{tabular}{l l l l}
		\hline 
		\textbf{UE array}& \textbf{4x4} & \textbf{4x4} & \textbf{4x4} \\
		\hline 
		\textbf{BS array}& \textbf{mmWave} & \textbf{THz A1} & \textbf{THz A2} \\
		\hline 
		\hline
		8x4 & 73.3 & - & -\\ 
		\hline
		16x4 & 91.3 & - & -\\
		\hline
		32x4 & 113.6 & - & -\\
		\hline
		64x4 & - & 20.1 & 92.1\\
		\hline
		128x4 & - & 25.7 & 114.4\\
		\hline
		256x4 & - & 32.6 & 142.2\\
		\hline
	\end{tabular} 
    \end{table}

\section{Numerical Analysis}\label{sec:num}

In this section, we numerically elaborate the association schemes and multi-connectivity strategies by utilizing the developed performance evaluation framework. We start our discussion with the effect of blockage on the new and ongoing session loss probabilities as well as mmWave BS resource utilization and then proceed with evaluating the impact of micromobility, and radio part parameters. The coverage radii of mmWave and THz BSs for the considered association schemes are presented in Table \ref{table:radii}, while the default  parameters utilized in our numerical study are summarized in Table \ref{table:numParam}. 

% Best guess on the chosen THz parameters

%Since there are no cellular or WLANs systems operating in THz band yet, the chosen set of systems parameters attempts to provide the “best guess” of what first generation of THz communications systems might come with. Particularly, we selected the carrier frequency that is least affected by atmospheric absorption and utilized in the only THz standard available to date, IEEE 802.15.3d \cite{petrov2020ieee}. The emitted power is chosen as a trade-off between the typical power utilized in cellular systems and potential restrictions of THz electronics in first generation THz communications systems. In spite the future THz system may have slightly different system parameters we believe that the analysis below sheds some light on the qualitative trade-offs involved in the system design.

% Fixing Cartesian movement

As shown in \cite{petrov2020capacity}, the effect of Cartesian displacements, $\Delta{x}$ and $\Delta{y}$, is negligible compared to yaw and pitch mobilities. Thus, in the rest of this section, we concentrate on the effect of the latter parameters and keep $\Delta{x}$ and $\Delta{y}$ constant at $3$ cm/s. Also, below we demonstrate results for on-demand beamalignment scheme and applications characterized by outage sensitivity $T_O=T_B$, where $T_B$ is the beamalignment time.

\subsection{Effect of Blockage}

\subsubsection{Outage Avoidance Association, A1}

% Association A1, blockage ongoing session loss pr.

\begin{figure*}[!t]
\vspace{-0mm}
\centering\hspace{-0mm}
\subfigure[{New session loss probability}]{
	\includegraphics[width=0.275\textwidth]{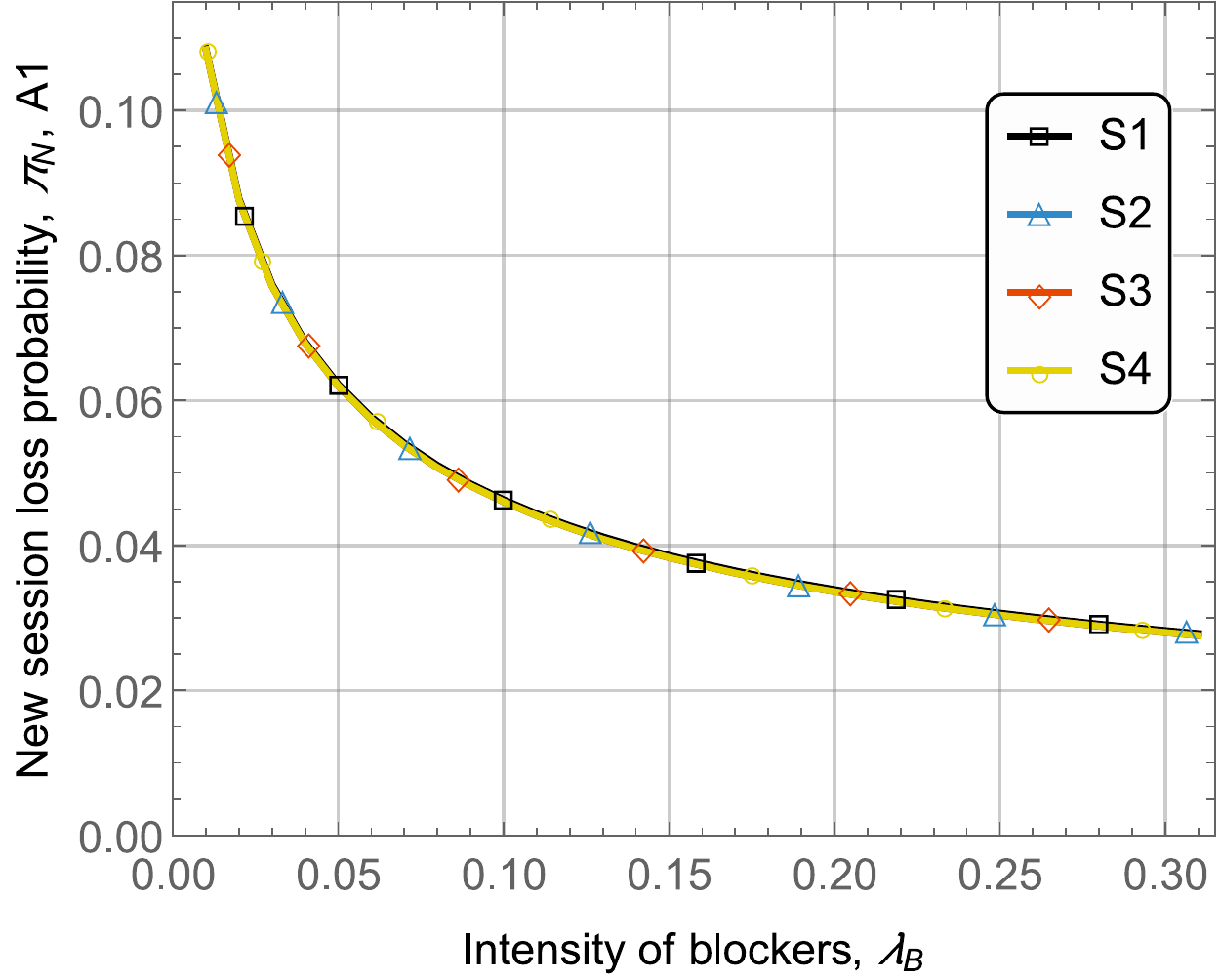}
	\label{fig:A1P1}
}~~~
\subfigure[{Ongoing session loss probability}]{
	\includegraphics[width=0.27\textwidth]{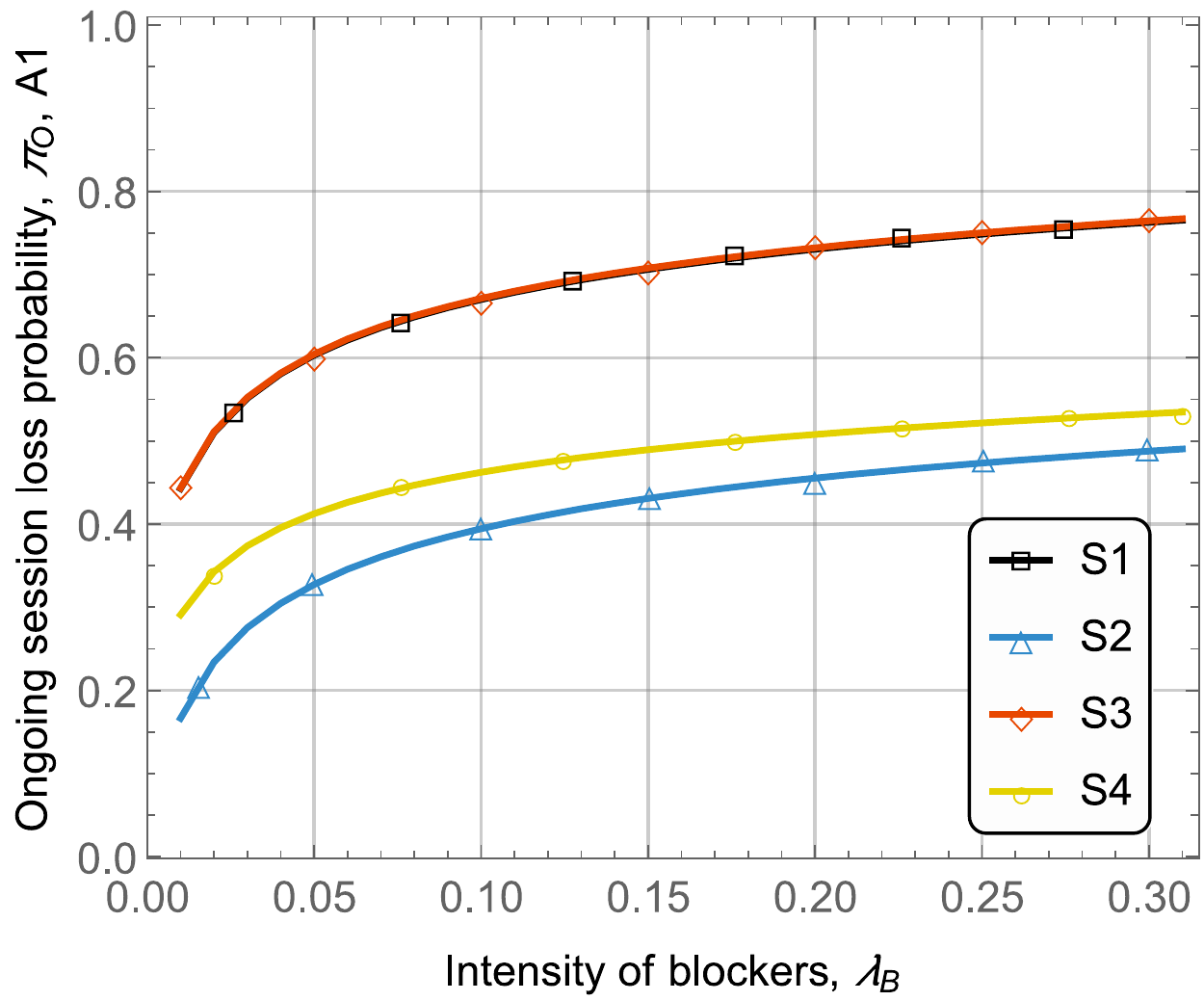}
	\label{fig:A1P2}
}~~~
\subfigure[{System resource utilization}]{
	\includegraphics[width=0.27\textwidth]{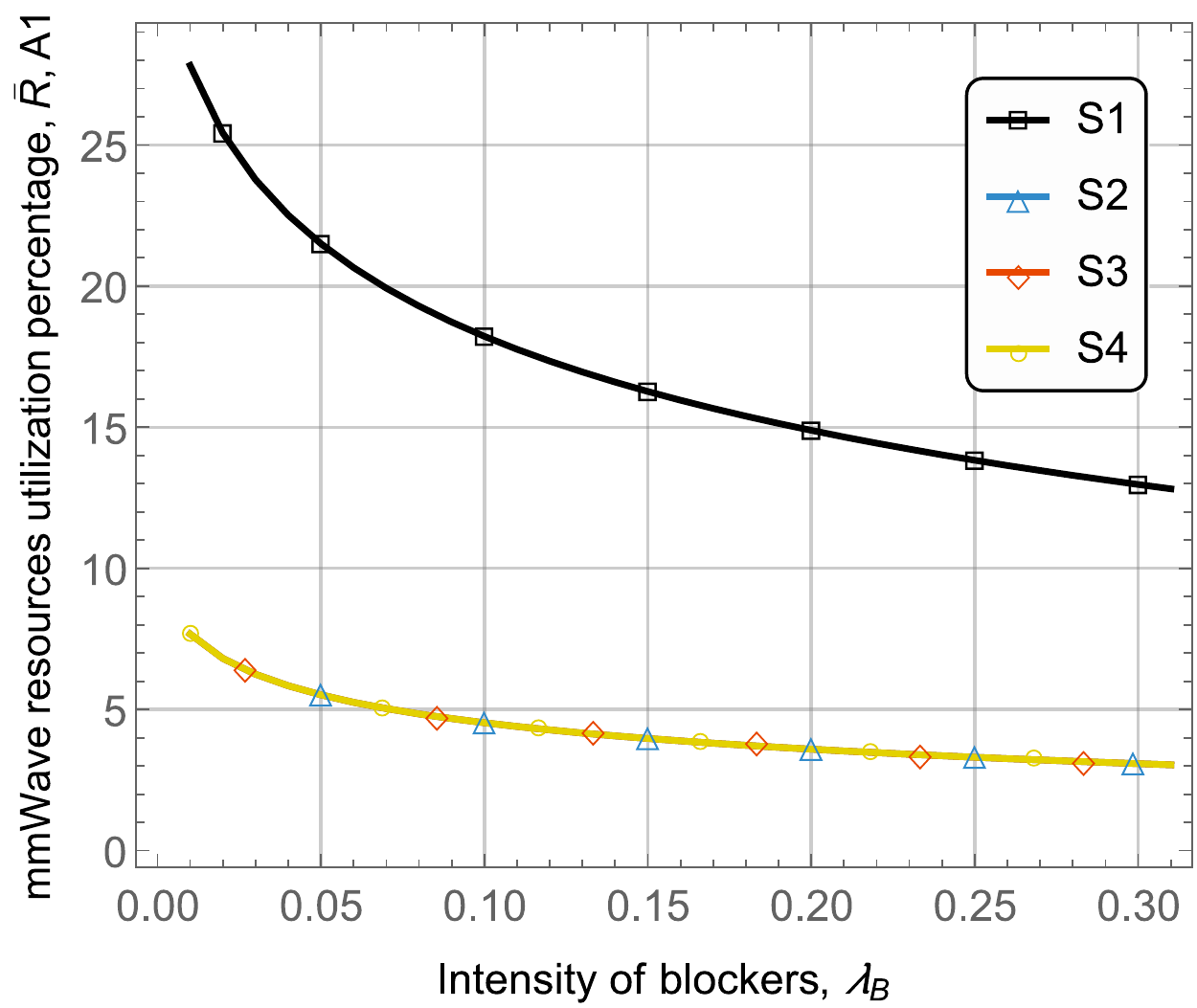}
	\label{fig:A1P3}
}
\caption{Considered performance metrics for A1 association scheme and different multi-connectivity strategies.}
\label{fig:A1}
\vspace{-0mm}
\end{figure*}

We start our analysis with comparison of multi-connectivity strategies for A1 association scheme, where UEs are associated with THz BSs only when blockage does not lead to outage. To this aim, Fig. \ref{fig:A1} illustrates new and ongoing session loss probabilities as well as system resource utilization for different multi-connectivity strategies, S1-S4, as a function of the blockers intensity, $C=10$ Mbps, $\Delta\phi=\Delta\theta=0.1^\circ$/s, $8\times{4}$ and $64\times{}4$ mmWave and THz BS antenna arrays, respectively, the number of THz BS in the coverage of mmWave BS $N=5$, and session arrival intensity of $10^{-4}$ sessions/s/m$^2$. Recall, that for the considered association scheme blockage does not affect the service process of sessions that are initially associated with THz BSs. However, it does cause reallocation of resources for sessions at mmWave BS. Furthermore, sessions originally associated with THz BSs are affected by the micromobility. The latter two effects cause the increase in the ongoing session loss probability for all the considered multi-connectivity strategies as evident from Fig. \ref{fig:A1P2}. However, the underlying reason is different for S1/S3 multi-connectivity strategies as compared to S2/S4. Recall, that for S1/S3 micromobility always leads to the loss of sessions. For this reason, the ongoing session loss probability coincides for these schemes. Further, S2 presumes the use of outage non-sensitive applications that can tolerate outage caused by antenna misalignment and thus the value on the considered parameter is lowest out of all the considered strategies. Notice that it is still not negligible as reallocation of the resources for sessions due to blockage at mmWave BS leads to consistent increase in this metric as blockers' intensity increases. Finally, as S4 assumes that micromobility always causes rerouting from THz BS to mmWave BS it leads to "intermediate" values of the ongoing session loss probabilities as some of the sessions switched over to mmWave BS are lost due to insufficient available resources.

\begin{figure*}[!t]
\vspace{-0mm}
\centering\hspace{-0mm}
\subfigure[{New session loss probability}]{
	\includegraphics[width=0.275\textwidth]{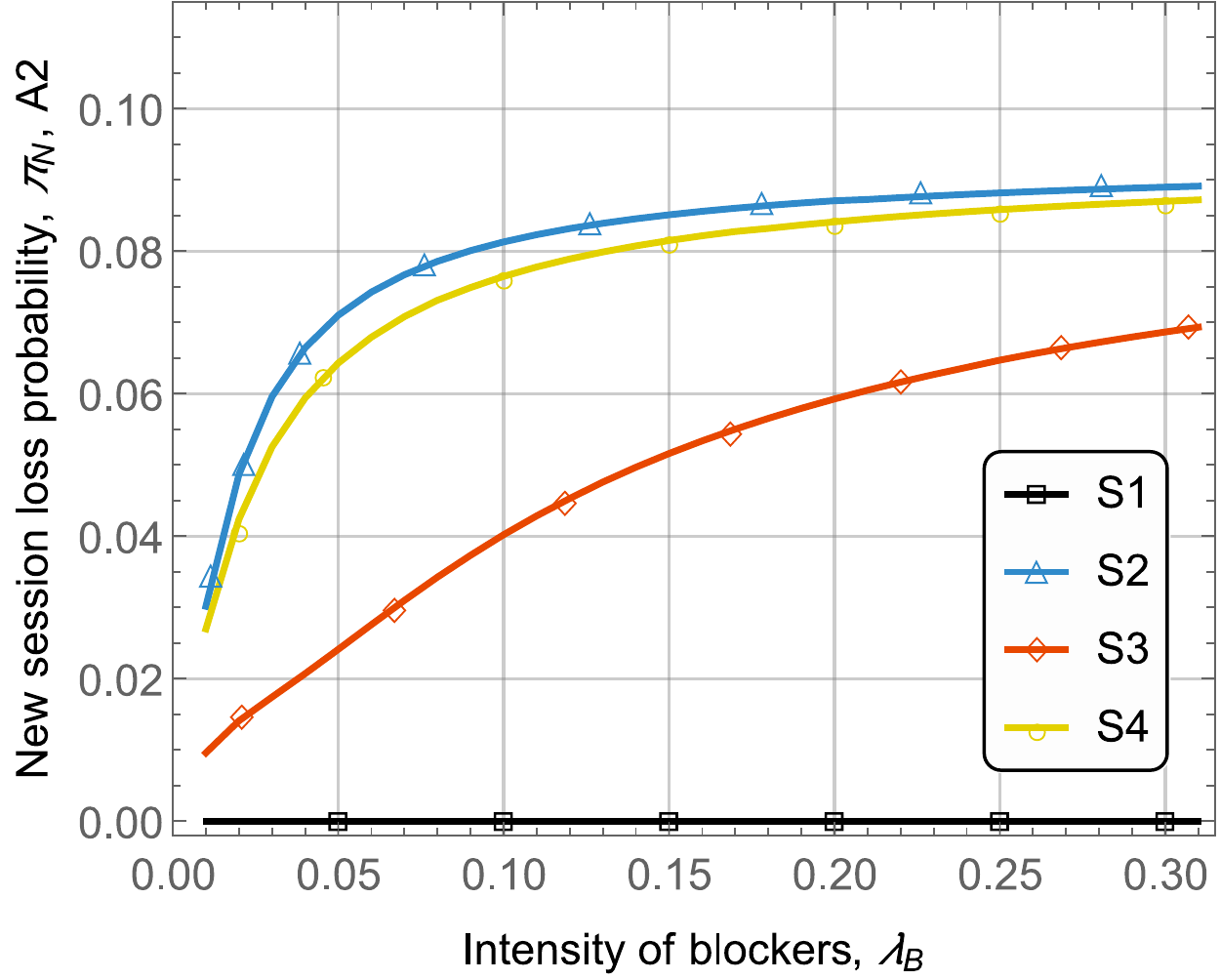}
	\label{fig:A2P1}
}~~~
\subfigure[{Ongoing session loss probability}]{
	\includegraphics[width=0.27\textwidth]{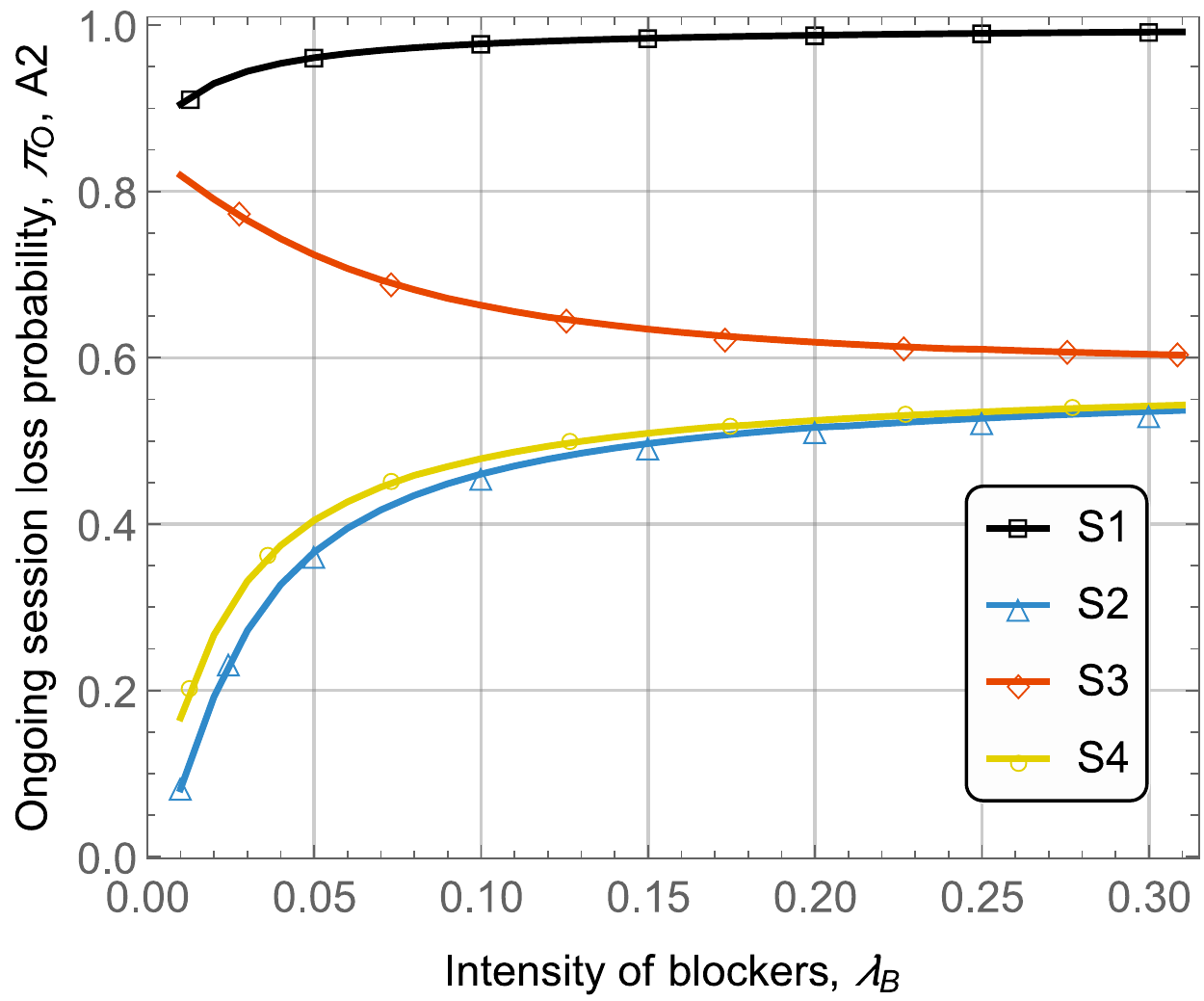}
	\label{fig:A2P2}
}~~~
\subfigure[{System resource utilization}]{
	\includegraphics[width=0.27\textwidth]{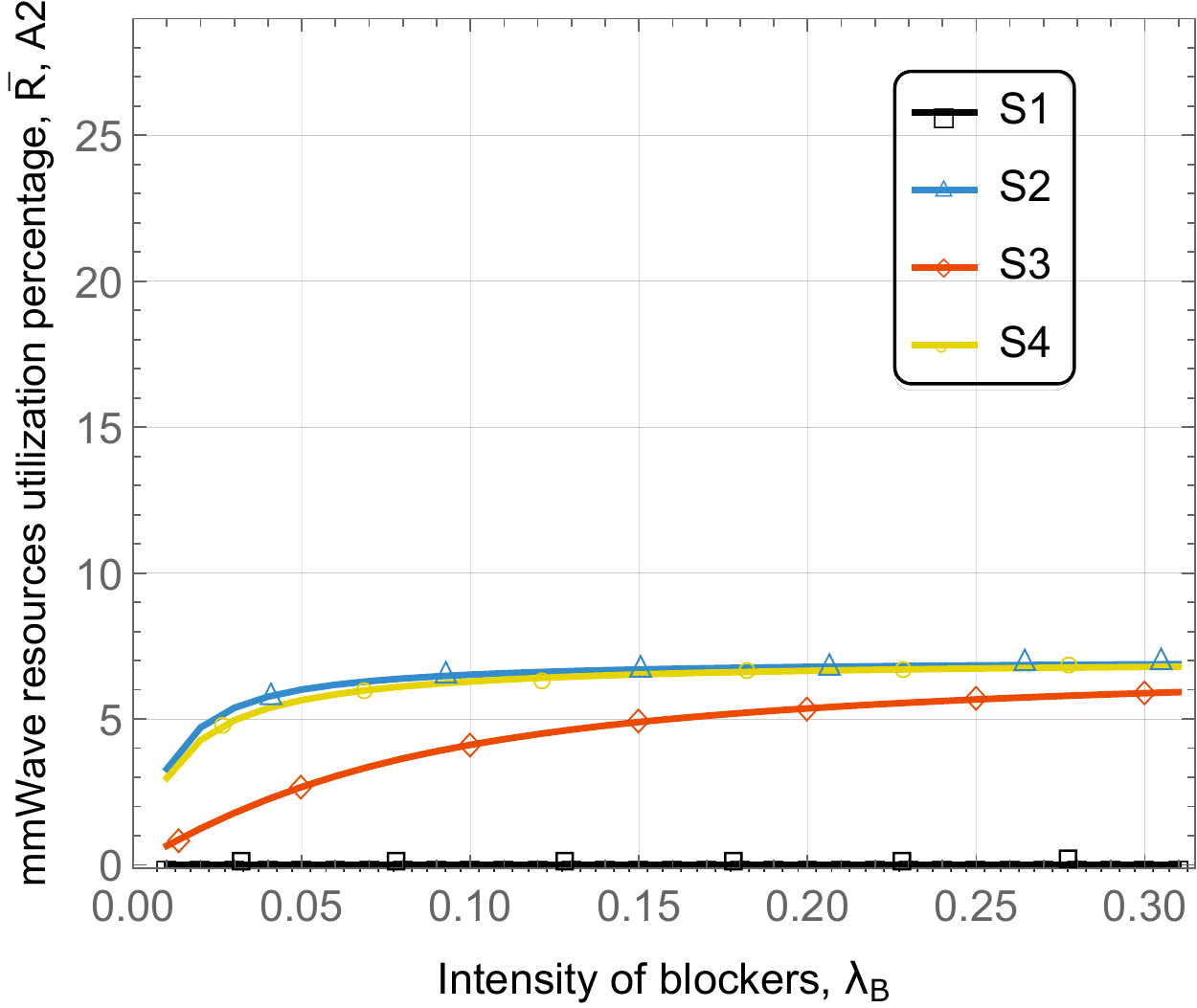}
	\label{fig:A2P3}
}
\caption{Considered performance metrics for A2 association scheme and different multi-connectivity strategies.}
\label{fig:A2}
\vspace{-4mm}
\end{figure*}

% New session loss pr.

Observe that Fig. \ref{fig:A1P1} shows that the new session loss probabilities coincide for all the considered multi-connectivity strategies. The rationale is that for A1 association scheme only micromobility may cause additional load at mmWave BS as sessions that may experience outage at THz BSs are accepted to mmWave BS. However, the time to align the antenna beams is very small producing negligible load at mmWave BS leading to the same performance for all the considered multi-connectivity strategies.

% Trade-off and resource utilization

Analyzing the data presented in Fig. \ref{fig:A1} further, one may observe that there is a trade-off between new and ongoing session loss probabilities for all the considered multi-connectivity strategies. Specifically, as the former metric decreases 
with blockers' intensity, the latter -- increases. The reason is that the increase in the ongoing session loss probability leads to more resources available for new sessions. These conclusions are also supported by the mmWave BS resource utilization which decreases with the blockage intensity $\lambda_B$. Observe that for the same new and ongoing session loss probabilities smaller resource utilization of mmWave BS implies overall better offloading performance of the system as more sessions are served at THz BS. By analyzing the data presented in Fig. \ref{fig:A1P3} one may conclude that S2/S4 strategies show better THz BSs resource utilization compared to S1/S3 strategies.

%However, by accounting for the resource utilization of mmWave BS resources demonstrated in Fig. \ref{fig:A1P3}, we may observe that the overall effect of blockage is still negative as it decreases with $\lambda_B$. Note that the system shows better utilization for S1/S3 multi-connectivity schemes as compared to S2/s4 ones.

\subsubsection{Coverage Enhancement Association, A2}

% Association A2, blockage

Having studied the association scheme A1, where a session that may experience outage in case of blockage at THz BSs is routed to mmWave BS, we now proceed assessing A2 association scheme, where these sessions are accepted at THz BSs. To this aim, Fig. \ref{fig:A2} shows new and ongoing session loss probabilities as well as system resource utilization for different multi-connectivity strategies and the same parameters as in Fig. \ref{fig:A1}. Here, one may observe that the qualitative behavior of the metrics is reversed. More specifically, the new session loss probability increases as opposed to the decrease for A1. While the rationale for this behavior is generally attributed to blockage at THz BSs that leads to handing sessions over to the mmWave BS, there are specifics for each considered multi-connectivity strategy.

% S1 and S2

First of all, the simplest multi-connectivity strategy that showed relatively good performance for A1 association scheme, S1, where no rerouting is performed in case of outage at THz BS, now is characterized by the absolute worst performance. More specifically, for A2 association scheme, this strategy accepts almost all the sessions to the THz BSs but all of those are eventually lost due to blockage and micromobility. In contrast, the multi-connectivity strategy S4, where both micromobility and outage at THz BSs causes change of the association point to mmWave BS, is characterized by similar ongoing session loss probability to A1 association scheme but has an increased new session loss probability. However, notice that in the negligible blockers density regime, i.e., $\lambda_B<0.1$ bl./m$^2$, S4 multi-connectivity strategy outperforms any other strategy for A1/A2 association schemes and can be generally recommended for network operators. 

% S2 (no micromobility)/S3(loss due to micromob)

Logically, S2 multi-connectivity strategy, characterizing performance of applications non-sensitive to short-term outage caused by micromobility, is associated with slightly better ongoing session loss probability but worse new session loss probability. For A2 association scheme unique strategy is S3, where no actions is taken against micromobility resulting in session losses, but blockage is avoided by handing the session over to mmWave BS. For this strategy, the ongoing session loss probability decreases with blockers density but the associated increase in the new session loss probability is milder compared to S2 and S4 strategies. The rationale is that with the increase of the blockers density $\lambda_B$, the probability that the session is rerouted to mmWave BS increases diminishing the chances of session being dropped due to micromobility at THz BSs. In spite A2/S3 combo still loses to A1/S4, it might be a viable option for future THz/mmWave deployments.

% some general concluding remarks

Thus, we may conclude that accepting sessions to THz BSs that may experience outage in case of blockage is generally worse as compared to serving them at mmWave BS for non-negligible blockers density. Contrarily, for $\lambda_B<0.1$ the combination of A1 association scheme and S2/S4 multi-connectivity strategies show the best performance.

\subsubsection{The Ongoing Session Loss Probability}

\begin{figure}[!t]
\vspace{-0mm}
\centering\hspace{-0mm}
\subfigure[{Association scheme A1}]{
	\includegraphics[width=0.3\textwidth]{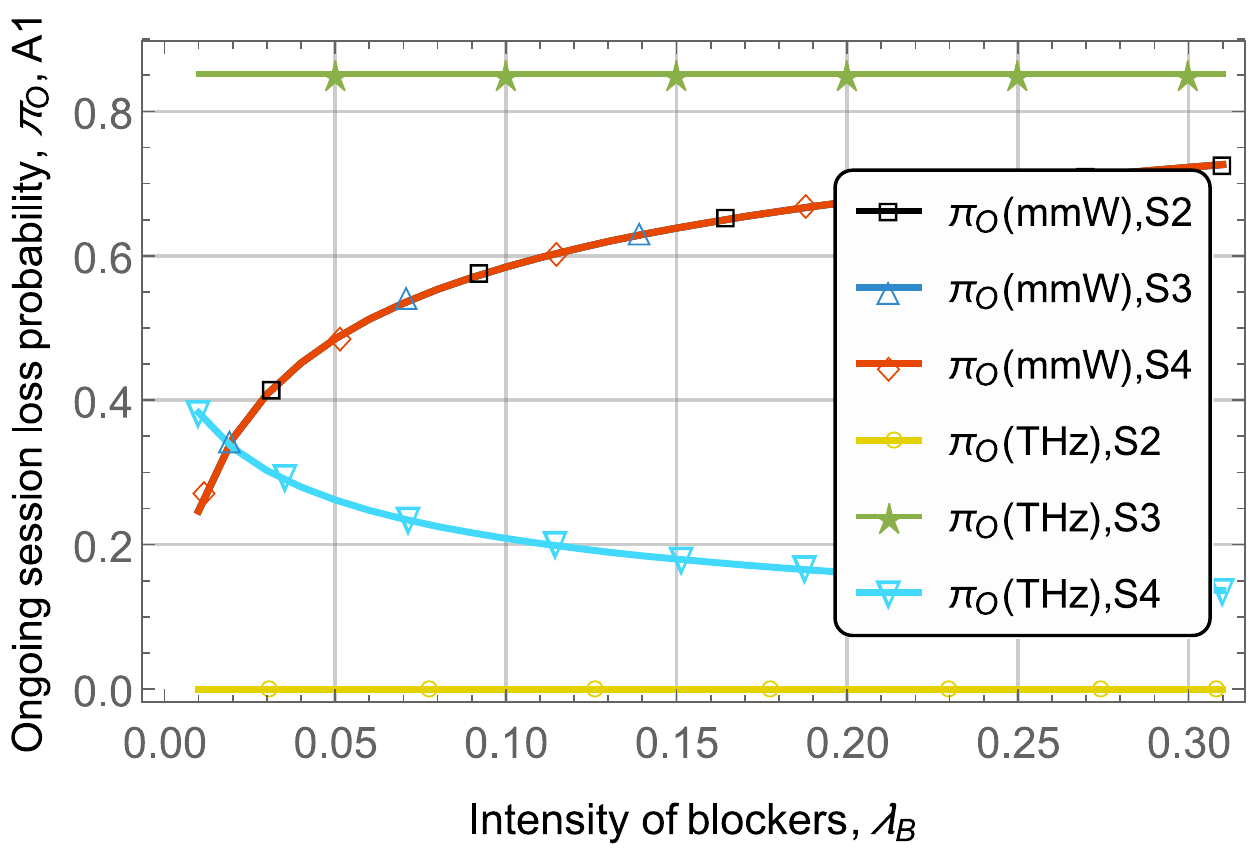}
	\label{fig:compA1}
}\\
\subfigure[{Association scheme A2}]{
	\includegraphics[width=0.3\textwidth]{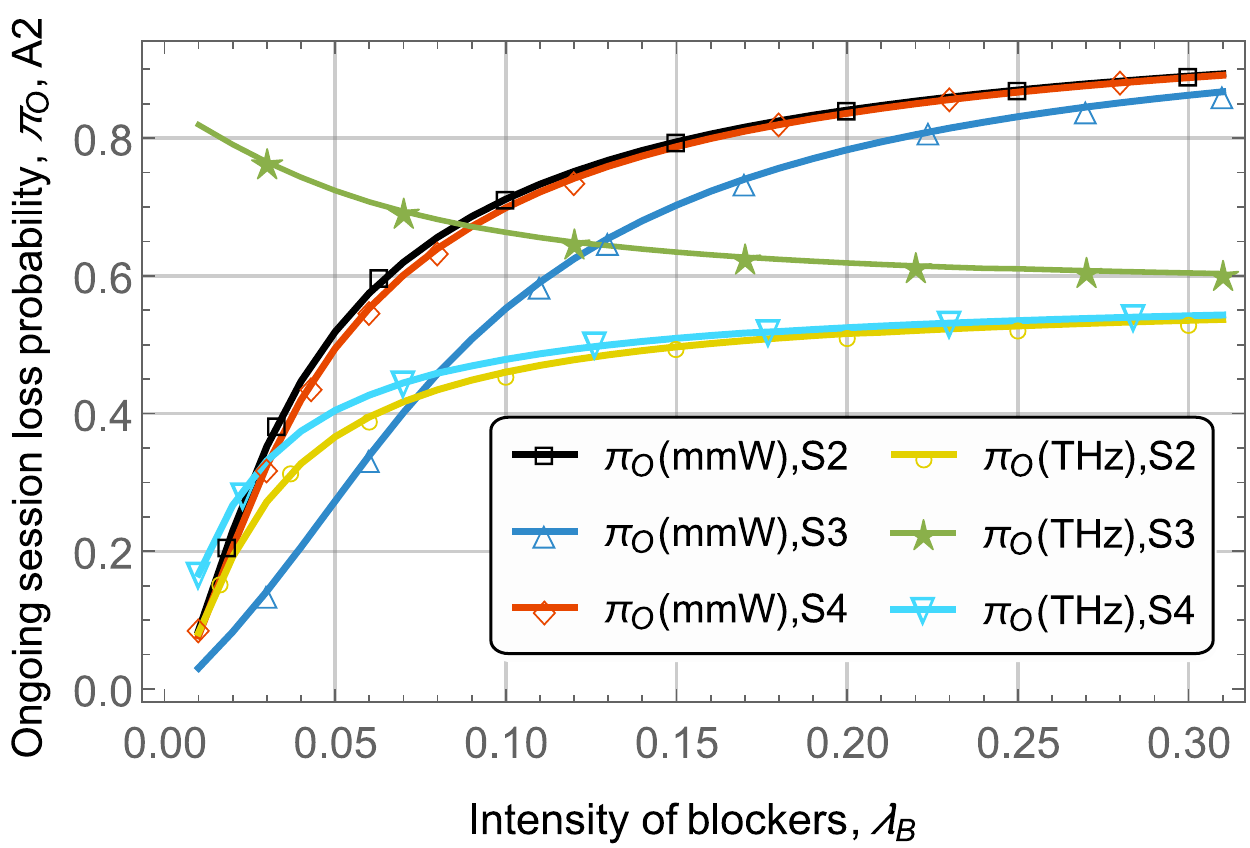}
	\label{fig:compA2}
}
\caption{Components contributing to the ongoing session loss probability.}
\label{fig:components}
\vspace{-4mm}
\end{figure}

% Two factors contributing

The considered multi-connectivity strategies, S2-S4, are characterized by two components contributing to the ongoing session loss probability: (i) losses at mmWave BS as a result of LoS blockage and subsequent unsuccessful attempt of resource reallocation and, (ii) losses as a result of rerouting from THz BSs to mmWave BS. To understand the structure of the considered metric, Fig. \ref{fig:components} presents these components as a function of the blockers' density for the same system parameters as in Fig. \ref{fig:A1}, Fig. \ref{fig:A2}.

% A1 strategy

Analyzing the presented data, one may observe that for the association scheme A1, the ongoing session loss probability at the mmWave BS coincide for all the considered schemes and increase with the blockers' density. The main reason is reallocation of resources caused by sessions currently served at mmWave BS and switching from non-blocked to blocked states. The component associated with the session loss at THz BS is constant at zero for outage non-sensitive applications (S2 strategy) and is maximized for strategy S3, where no actions is taken in case of micromobility. Finally, the strategy S4 is characterized by the decreasing loss probability at THz BSs. Still, the increase caused by session loss at mmWave BS prevails leading to the associated increase in Fig. \ref{fig:A1P2}.

% A2 strategy

For A2 association scheme, where the session is accepted to THz BSs even when blockage leads to outage, mmWave BS loss components are all increasing functions of $\lambda_B$. However, the behavior of session loss probability associated with THz BS is different. The reason is the interplay between blockage and micromobility processes at THz BS. For S2 multi-connectivity strategy, where outage non-sensitive applications capable of surviving beamalignment caused by micromobility are assumed, the ongoing session loss probability at THz BS increases. The reason is session reroutes to mmWave BS caused by blockage. However, for S3 multi-connectivity strategy, where micromobility leads to the loss of sessions the increase in the blockers' density increases the ongoing session loss probability at THz BS. This component dominates leading to the trends observed in Fig. \ref{fig:A2P2}.

\subsection{Effect of Micromobility}

\begin{figure}[!t]
\vspace{-0mm}
\centering\hspace{-0mm}
\subfigure[{New session loss probability}]{
	\includegraphics[width=0.3\textwidth]{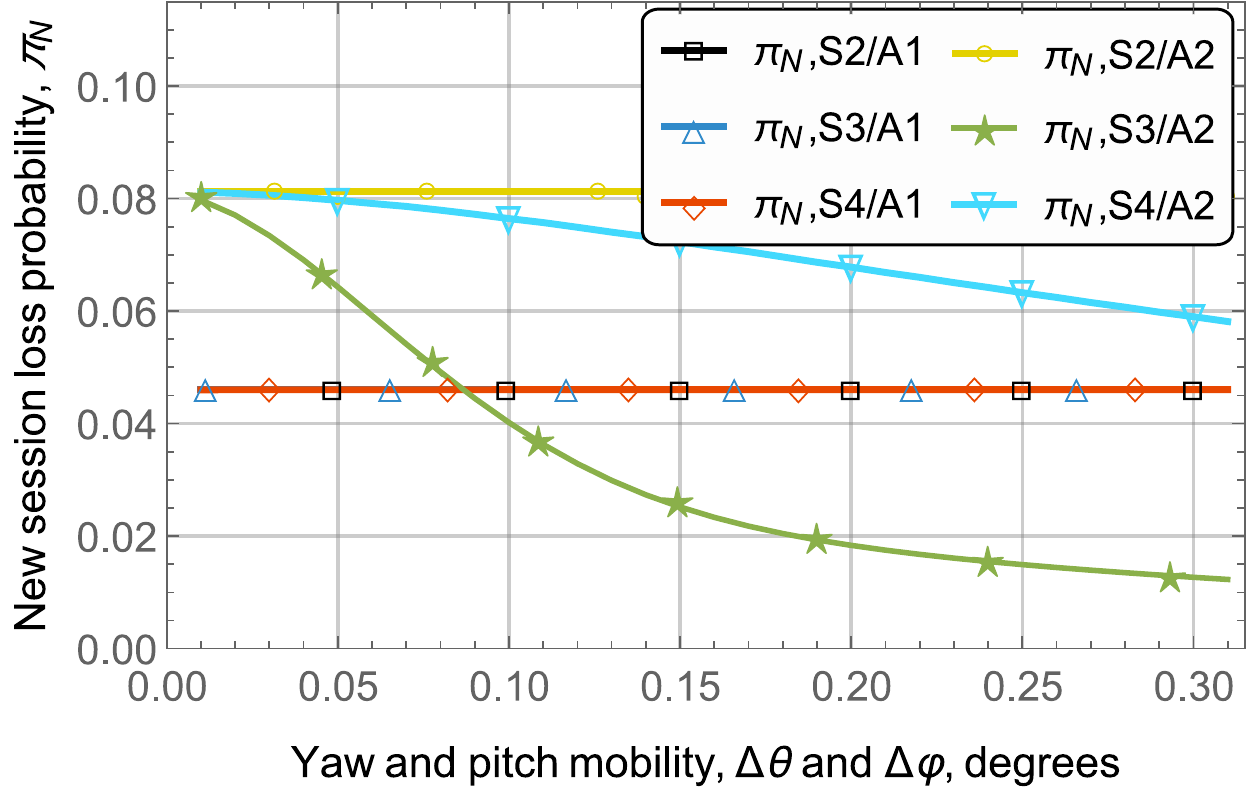}
	\label{fig:microNew}
}\\
\subfigure[{Ongoing session loss probability}]{
	\includegraphics[width=0.3\textwidth]{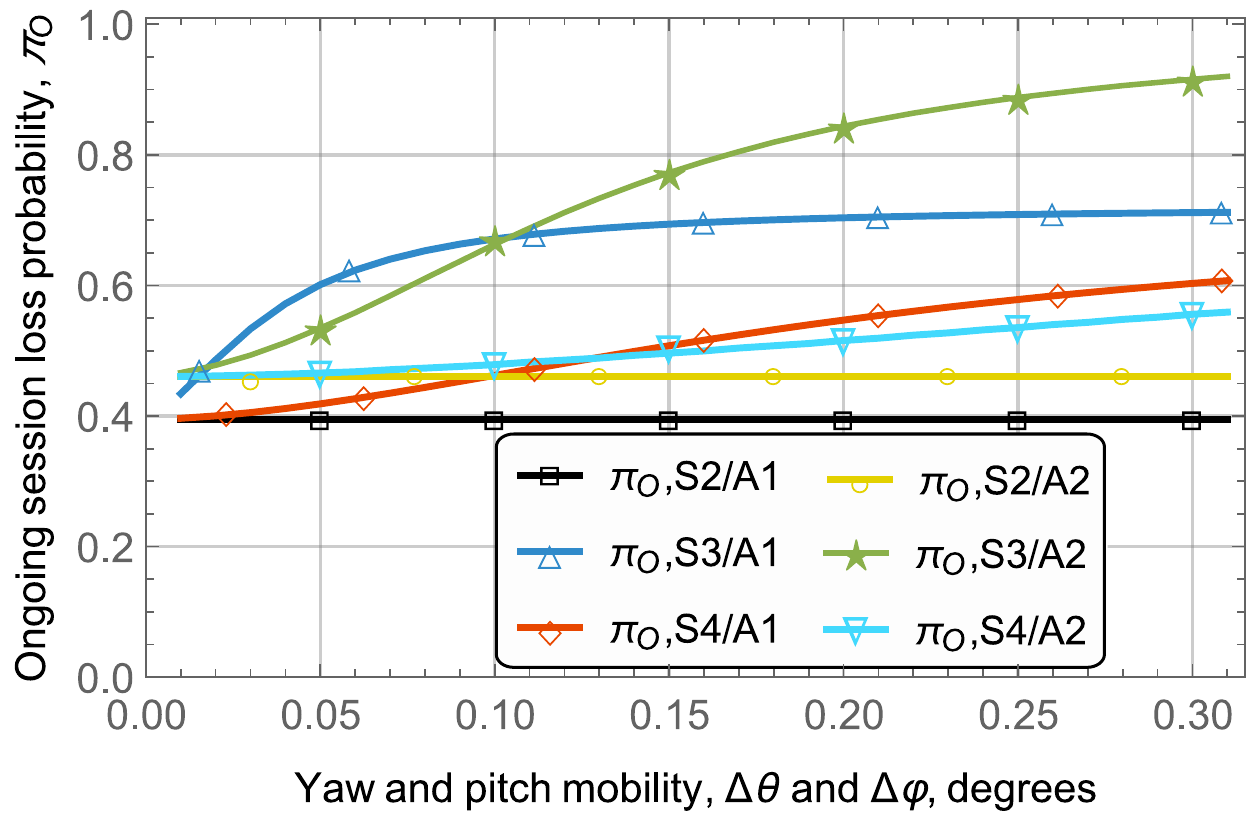}
	\label{fig:microOng}
}
\caption{The impact of micromobility on session loss probabilities.}
\label{fig:micro}
\vspace{-4mm}
\end{figure}

As we have already observed, the presence of micromobility provides significant impact of the system performance. To investigate the extent of this impact we now study the response of the new and ongoing session loss probabilities to the yaw and pitch mobilities, $\Delta\phi=\Delta\theta$, in Fig. \ref{fig:micro} for $C=10$ Mbps, blockers density of $\lambda_B=0.1$ bl./m$^2$, $8\times{4}$ and $64\times{}4$ mmWave and THz BS antenna arrays, respectively, the number of THz BS in the coverage of mmWave BS $N=5$, and session arrival intensity of $10^{-4}$ sessions/s/m$^2$.

% A1 association scheme

By analyzing the data presented in Fig. \ref{fig:micro}, one may observe that all qualitative trends remain intact over the considered range of micromobility speeds. For A1 association scheme, where sessions are only accepted to THz BSs when blockage does not lead to outage, the new session loss probabilities remain unaffected by the micromobility speed. Specifically, micromobility does not affect S2 strategy at all, as the application is assumed to tolerate beamalignment time. For S3 and S4 strategies, micromobility speed does impact ongoing session loss probability and in the case of S4 also the load imposed at mmWave BS. However, the effect of the beamalignment time is negligible producing almost no impact on the mmWave BS load and thus the new session loss probability. 

% A2 association scheme

The situation is drastically different for A2 association scheme, where a session can be accepted to THz BS even when blockage may lead to outage conditions. Here, we see the significant decrease in the new session loss probabilities as a result of reduced load imposed on mmWave BS due to significantly higher ongoing session losses, see Fig. \ref{fig:microOng}. Noticeably, the ongoing session loss probability is impacted most by strategy S3 for both A1 and A2 association schemes, where no actions are taken in case of micromobility events. The S4 strategy allows to effectively eliminate even extremely high micromobility speeds reaching $0.3^\circ$/s, while the strategy S2 is logically not affected at all. Concluding, we may state that the micromobility speed does not qualitatively change the conclusions stated above for association schemes and multi-connectivity strategies.

\subsection{Effects of the Traffic, Arrays, and Session rate}

\begin{figure}[t!]
\vspace{-0mm}
\centering\hspace{-0mm}
\subfigure[{New session loss probability}]{
	\includegraphics[width=0.3\textwidth]{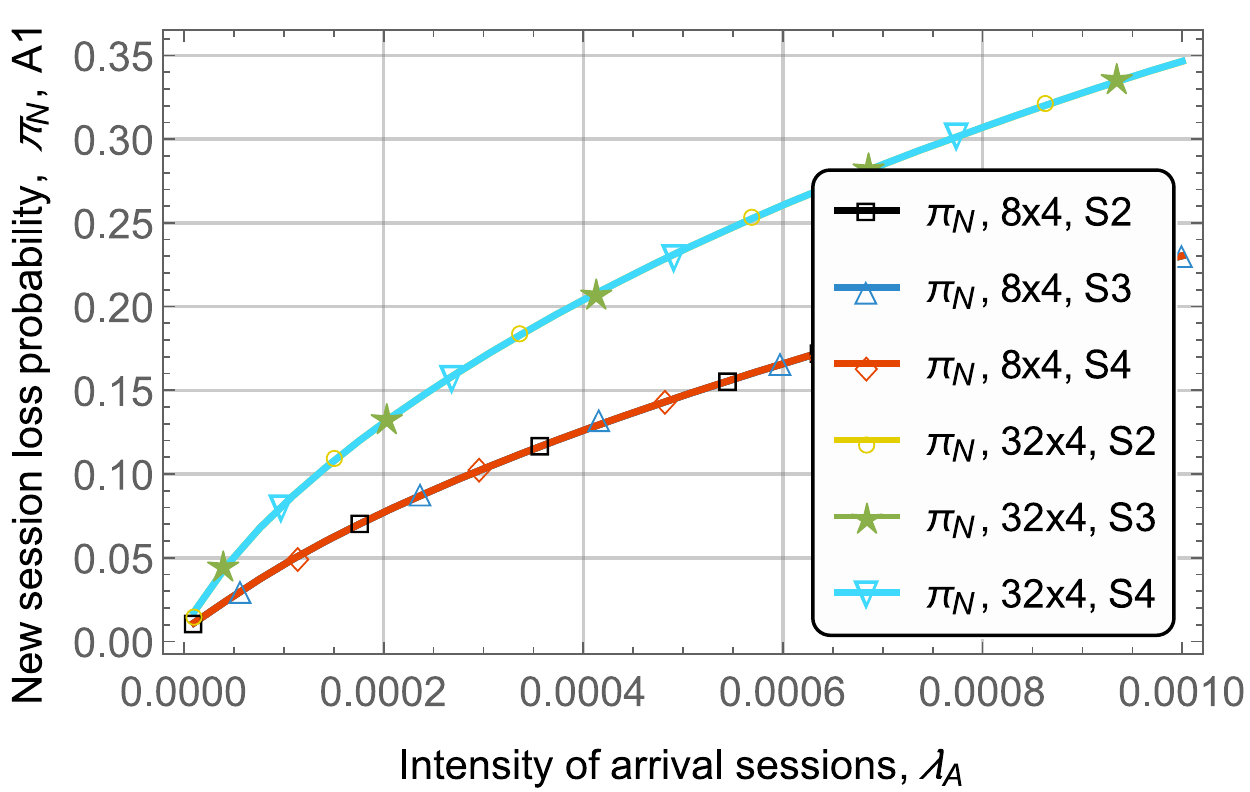}
	\label{fig:loadew}
}\\
\subfigure[{Ongoing session loss probability}]{
	\includegraphics[width=0.3\textwidth]{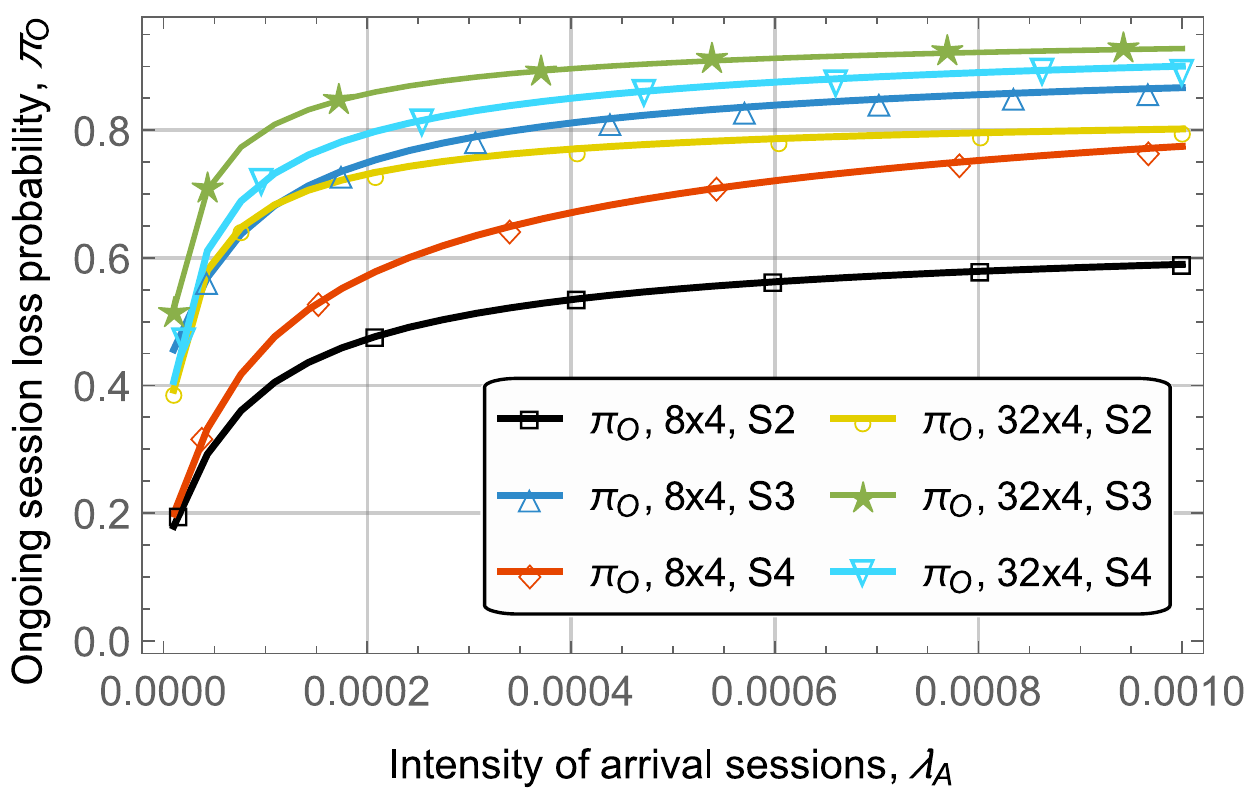}
	\label{fig:loadOng}
}
\caption{The impact of traffic load and antenna arrays.}
\label{fig:load}
\vspace{-4mm}
\end{figure}

\begin{figure}[b!]
\vspace{-4mm}
\centering\hspace{-0mm}
\subfigure[{As a function of blockers intensity}]{
	\includegraphics[width=0.3\textwidth]{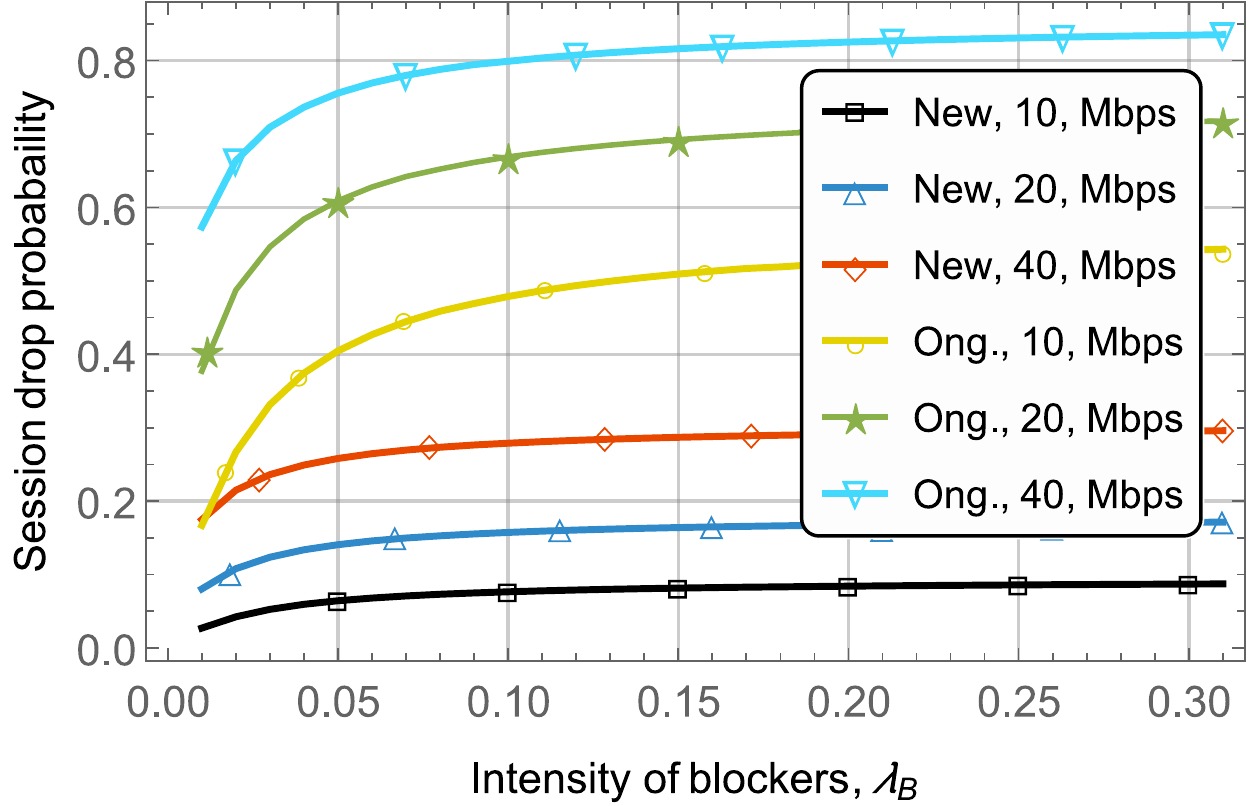}
	\label{fig:rate-blockers}
}\\
\subfigure[{As a function of micromobility}]{
	\includegraphics[width=0.3\textwidth]{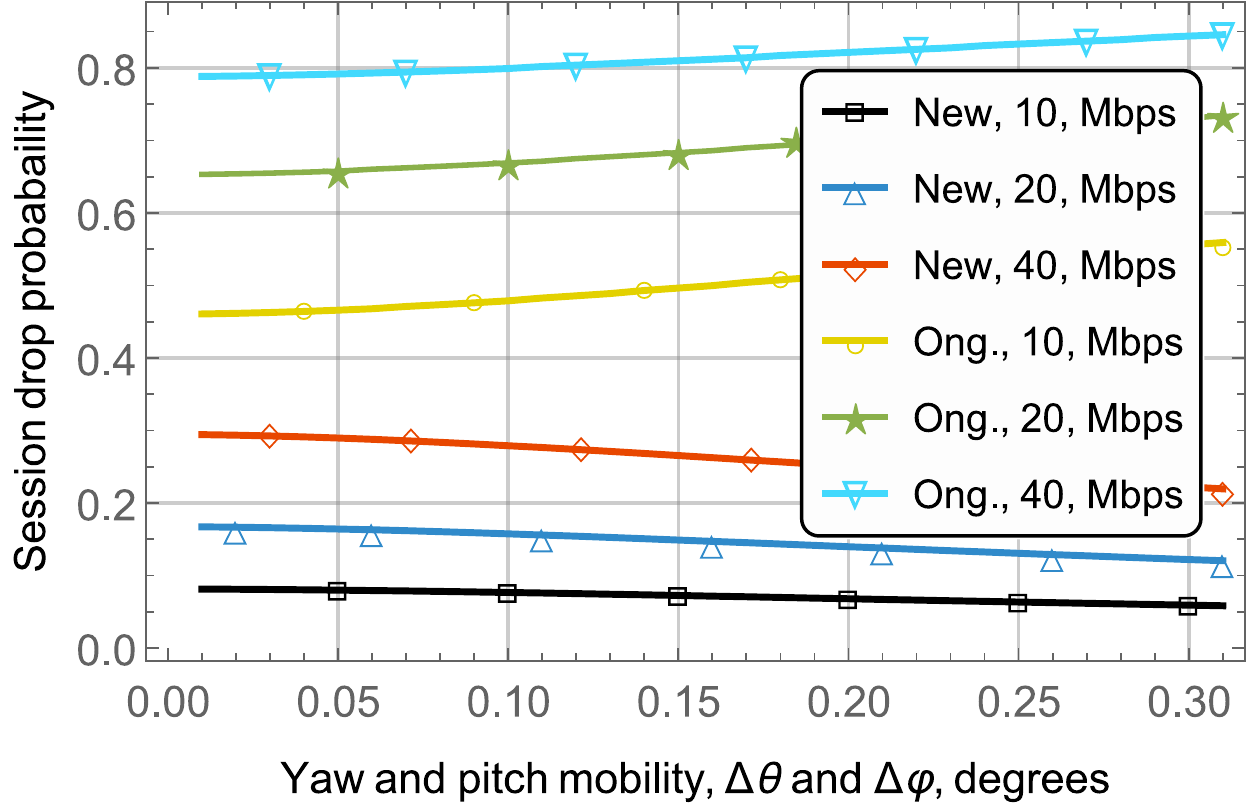}
	\label{fig:rate-micro}
}
\caption{The impact of session rate on drop probabilities.}
\label{fig:rateEffect}
\vspace{-0mm}
\end{figure}

We now evaluate whether session arrival intensity or radio part parameters affect the abovementioned conclusions. To this aim, Fig. \ref{fig:load} show the new and ongoing session loss probabilities as a function of the session arrival rate for A1 association scheme, two configurations of the mmWave BS antenna array, $8\times{}4$ and $32\times{}4$, $C=10$ Mbps, $\lambda_B=0.1$ bl./m$^2$, $\Delta\phi=\Delta\theta=0.1^\circ$/s, $64\times{}4$ THz BS antenna array, and $N=5$. 

As one may observe, the behavior of the new and ongoing session loss probabilities is preserved across the whole range of the offered traffic load. Logically, for a given antenna configuration, the new session loss probabilities coincide for all the considered multi-connectivity strategies. However, there is clear difference between performance associated with the utilized antenna arrays at mmWave BS. More specifically, the system with $8\times{}4$ arrays is characterized by significantly lower new and ongoing session loss probabilities as compared to the one with $32\times{}4$ array. The rationale for this behavior is however attributed to the densification property as the system with $32\times{}4$ array covers much larger area (see Table \ref{table:radii}) and is thus faced with higher traffic arrival intensity.

Finally, we consider the effect of the session rate. To this aim, Fig. \ref{fig:rateEffect} shows the new and ongoing session drop probabilities for different session rates as a function of blockers density A2 association and S4 mutliconnectivity schemes, $8\times{}4$ and $64\times{}4$ mmWave and THz antenna arrays, and $N=5$. As one may observe, the change in the session rates leads to quantitative and proportional change in the considered metrics.

\section{Conclusions}\label{sect:concl}

Motivated by the need to ensure service reliability of non-elastic rate-greedy applications in future 6G deployments featuring both mmWave and THz BS, we have analyzed the service process of such type of traffic in joint THz and mmWave cellular systems supporting multi-connectivity operation and subject to both blockage and micromobility impairments. To perform analysis of various user association schemes and multi-connectivity strategies, we have developed a unified mathematical framework simultaneously capturing radio and service specifics at the THz and mmWave BSs. 

Our numerical analysis demonstrates that, when utilizing multi-connectivity at UEs, for high blockers density environment, i.e., $\lambda_B>0.1$ bl./m$^2$, only those sessions that does not experience outage in case of blockage, should be accepted for service at THz BSs. Otherwise, the coverage range of THz BSs can be enlarged by accepting also those sessions that may experience outage. Out of all the considered multi-connectivity strategies, the one that reroutes the session in case of both outage and multi-connectivity demonstrates consistently good results over the whole considered range of parameters. However, in this context, ensuring that the session may tolerate short-term outages caused by antenna misalignment at THz BSs is crucial as it greatly improves service reliability. Finally, under low blockage and micromobility conditions, the strategy that does not utilize multi-connectivity at all is characterized by the competitive performance. This is critical observation as it may simultaneously allow for energy conservation as supporting multi-connectivity is known to be energy consuming functionality \cite{li2020power}.

Comparing the obtained results to those provided in \cite{begishev2021performance} for “sub-6 GHz+mmWave” system we emphasize that “mmWave+THz” system is currently free of resource seizure effects, where the temporal offloading of mmWave sessions causes massive session drops at sub-6 GHz BSs. Thus, as opposed to the “mmWave+THz” systems considered in our study, the use of sub-6GHz BSs to serve mmWave sessions is only feasible in light traffic conditions. However, we stress that this conclusion remains valid for the current session rates in the order of 10-100 Mbps (that includes yet to be deployed applications such as XR/VR applications, holographic communications) as mmWave band can be utilized to temporally support multiple tens of that simultaneously.

\appendix

\begin{figure}[t!]
\centering
\includegraphics[width=.8\columnwidth]{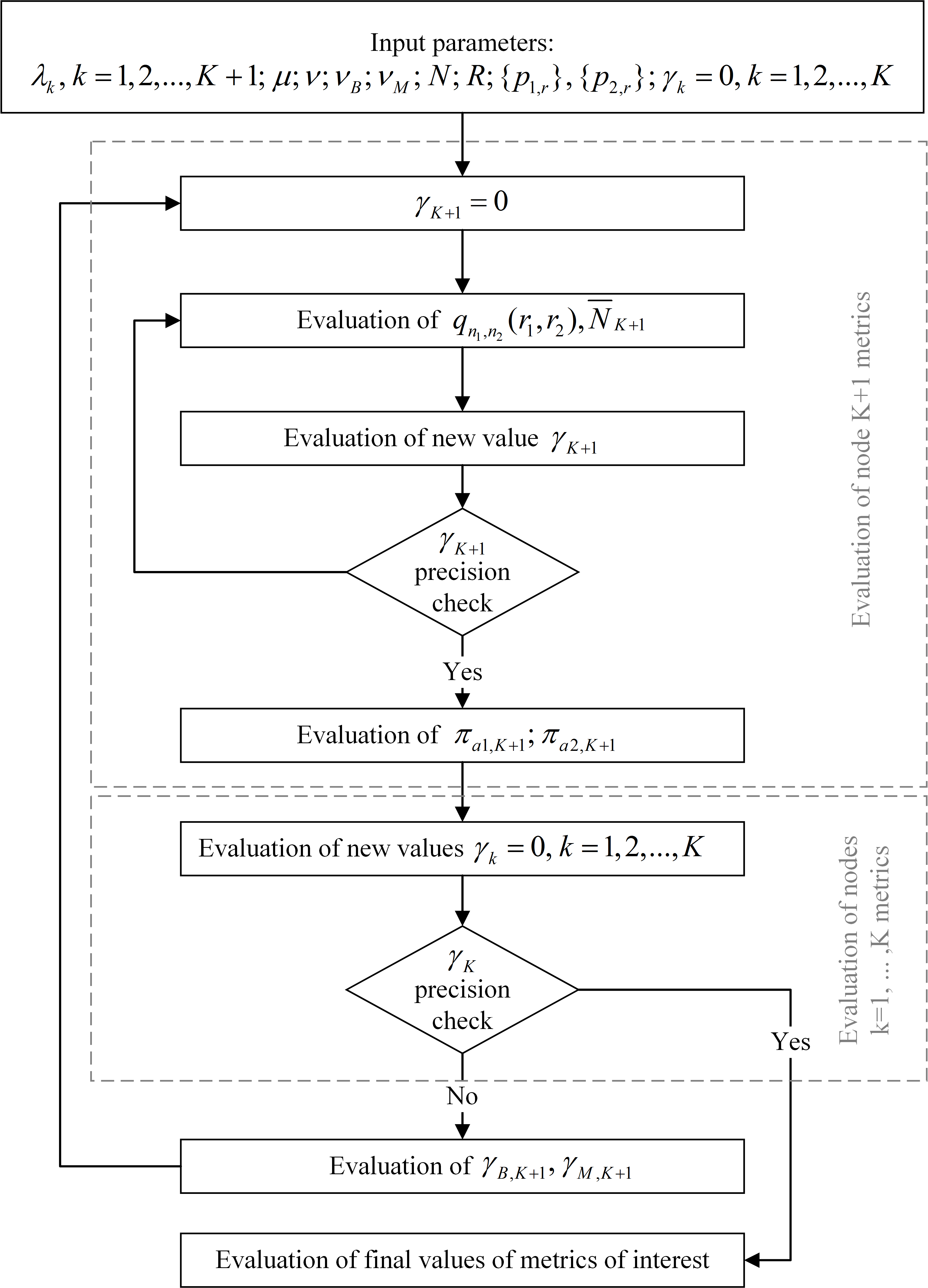}
\caption{Block diagram of the iterative algorithm.}
\label{fig:algorithm}
\vspace{-3mm}
\end{figure}

Calculation of (\ref{eq:gamma_k}) requires an iterative algorithm shown in Fig. \ref{fig:algorithm}. The algorithm starts with the evaluation of metrics at node $K+1$ with only primary session arrivals ($\gamma_{K+1}=\gamma_{B,K+1}=\gamma_{M,K+1}=0$). Stationary distribution $q_{n_1,n_2}(r_1,r_2)$, see \eqref{eq:q} and \eqref{eq:q0}, and average number of primary and secondary sessions $\bar{N}_{1,K+1}$ in \eqref{eq:barn12} are calculated. Using the obtained value of $\bar{N}_{1,K+1}$, the new value of the secondary sessions arrival intensity $\gamma_{K+1}$ is calculated according to \eqref{eq:gammak}. If the modulo of the difference between the new and the previous values of $\gamma_{K+1}$ is greater than the predefined precision level, the algorithm proceeds to the new iteration. If the desired precision is achieved, the session loss probabilities $\pi_{a1,K+1}$ and $\pi_{a2,K+1}$ are obtained according to \eqref{eq:pia1}. Besides, the new values of $\gamma_k$ are evaluated using \eqref{eq:gamma_k}. Again, if the modulo of the difference between the new and the previous values of $\gamma_k$ is greater than the precision level, the arrival intensities $\gamma_{B,K+1}$, $\gamma_{M,K+1}$ of the rerouted sessions at node $K+1$ are calculated according to \eqref{eq:gammabk}, and the algorithm returns to the evaluation of stationary probabilities with new arrival intensities of rerouted sessions and no secondary sessions, i.e. $\gamma_{K+1}=0$. Otherwise, the stable solution is achieved, and the algorithm proceeds to calculation of final metrics. %The block-scheme of the algorithm is depicted in Fig. \ref{fig:algorithm}.

\balance
\bibliographystyle{ieeetr}
\bibliography{UserAss}

\end{document}